# Mutant reduction evaluation: what is there and what is missing?

Peng Zhang   Yang Wang   Xutong Liu   Yanhui Li*   Yibao Yang

State Key Laboratory for Novel Software Technology, Nanjing University, China
Department of Computer Science and Technology, Nanjing University, China

Ziyuan Wang
School of Computer, Nanjing University of Posts and Telecommunication, China

Xiaoyu Zhou
School of Computer Science and Engineering, Southeast University, China

Lin Chen   Yuming Zhou*
State Key Laboratory for Novel Software Technology, Nanjing University, China
Department of Computer Science and Technology, Nanjing University, China

*Corresponding author

**Background.** Mutation testing is a commonly used fault injection technique for evaluating the effectiveness of a test suite. However, it is usually computationally expensive. Therefore, many mutation reduction strategies, which aim to reduce the number of mutants, have been proposed.

**Problem.** It is important to measure the ability of a mutation reduction strategy to maintain test suite effectiveness evaluation. However, existing evaluation indicators are unable to measure the "order-preserving ability", i.e., to what extent the mutation score order among test suites is maintained before and after mutation reduction. As a result, misleading conclusions can be achieved when using existing indicators to evaluate the reduction effectiveness.

**Objective.** We aim to propose evaluation indicators to measure the "order-preserving ability" of a mutation reduction strategy, which is important but missing in our community.

**Method.** Given a test suite on a Software Under Test (SUT) with a set of original mutants, we leverage the test suite to generate a group of test suites that have a partial order relationship in fault detecting potential. When evaluating a reduction strategy, we first construct two partial order relationships among the generated test suites in terms of mutation score, one with the original mutants and another with the reduced mutants. Then, we measure the extent to which the two partial order relationships are consistent. The more consistent the two partial order relationships are, the stronger the Order Preservation (*OP*) of the mutation reduction strategy is, and the more effective the reduction strategy is. Furthermore, we propose Effort-aware Relative Order Preservation (*EROP*) to measure how much gain a mutation reduction strategy can provide compared with a random reduction strategy.

**Result.** The experimental results show that OP and EROP are able to efficiently measure the "order-preserving ability" of a mutation reduction strategy. As a result, they have a better ability to distinguish various mutation reduction strategies compared with the existing evaluation indicators. In addition, we find that Subsuming Mutant Selection (SMS) and Clustering Mutant Selection (CMS) are more effective than the other strategies under OP and EROP.

**Conclusion.** We suggest, for the researchers, that OP and EROP should be used to measure the effectiveness of a mutant reduction strategy, and for the practitioners, that SMS and CMS should be given priority in practice.

# 1 INTRODUCTION

In software development, test suites are widely used to test a software system to examine whether it has faults. In this context, evaluating test effectiveness of a test suite is essential for many activities. On the one hand, it enables to know the fault detecting potential that a given test suite has (a fundamental question in software testing research [26]). Knowing such an "absolute" effectiveness score of a test suite can inform whether there is a need to generate new test cases to enhance its effectiveness. On the other hand, it enables to compare the fault detecting potential among different test suites (another fundamental question in software testing research [21]). Knowing the "relative" order of the test effectiveness score among test suites can provide a useful means for comparing the effectiveness of the corresponding test-generation techniques.

Currently, mutation testing is regarded as a useful technology to evaluate the effectiveness of a test suite. Specifically, the evaluation process proceeds as follows. First, a set of mutants are generated by modifying a given SUT (Software Under Test) in small ways (i.e. by applying a number of mutation operators that mimic typical programming errors). A mutant will be called "killed" if the test suite can distinguish the outputs between the SUT and the mutant; otherwise, it will be called "alive" under the test suite. Then, the effectiveness of the test suite is measured by mutation score, an indicator denoting the ratio of killed mutants to all the non-equivalent mutants (a mutant is called "equivalent" if its output cannot be distinguished from the output of the SUT by test cases). The higher the mutation score, the more effective the test suite is considered. In the literature, mutation score has been a commonly used indicator to evaluate the effectiveness of a given test suite. By mutation score, developers can not only know the "absolute" effectiveness of a given test suite, but also know the "relative" order of the test effectiveness among different test suites.

However, the main disadvantage of mutation testing comes from the huge cost of executing all mutants [1] [2]. Selecting a mutant subset from all mutants is a way to reduce overhead. In the last decades, many reduction strategies have been proposed based on two main practices. The first practice is leveraging the characteristics of the mutants to select the mutants with an execution value (e.g. selecting the mutants generated by specific mutation operators [3-7]). The rationale behind this practice is that some mutants are more relevant to faults. As a result, it is more practical to leverage their characteristics to search the subset. The second practice is selecting a mutant subset to maximum a pre-defined optimization function [8]. One of the most used optimization functions is to minimize the variation of mutation scores. In other words, this practice usually attempts to search the mutant subset with an approximate mutation score to the original mutation score, against a given test suite. The rationale behind setting such an objective function is that the criterion of mutant reduction is to keep the evaluation of a test set consistent, which is approximated by mutation score.

Given a mutant reduction strategy, an important question is how to objectively evaluate its reduction effectiveness, i.e., the ability to maintain test suite effectiveness evaluation. In nature, mutation score is used as a metric to measure the effectiveness of test suites. A mutant reduction strategy aims to provide a simple way to compute a new mutation score as a new metric to measure the effectiveness of test suites. In this context, we should measure the difference between the two metrics to evaluate the effectiveness of the mutant reduction strategy. Specifically, under the original mutants, developers can obtain the "absolute" effectiveness of each single test suite as well as the "relative" order of the effectiveness among different test suites. Ideally, a mutant reduction strategy should not lead to a change in the "absolute" effectiveness of each single test suite as well as the "relative" order of the effectiveness among different test suites. In our community, it is popular to employ the following process to evaluate the effectiveness of a mutant reduction strategy [4,6-11]. First, select a project with a sufficient test suite as the subject SUT. Second, execute the mutation operators on the SUT to generate a set of mutants. Third, apply a reduction strategy to the generated mutants to obtain a set of reduced mutants. Fourth, run the sufficient test suite against the original and reduced mutant sets to obtain two mutation scores. Fifth, use the Variation of two Mutation Scores (*VMS*) as an indicator to evaluate the effectiveness of the reduction strategy: the smaller the *VMS* is, the better the reduction strategy is considered. As can be seen, for a mutant reduction strategy, the above process uses *VMS* to evaluate its ability to maintain the "absolute" effectiveness of a single test suite (measured by the mutation score corresponding to the original mutants). However, this evaluation process ignores measuring its ability to maintain the "relative" order of the effectiveness among test suites. As a result, it is not possible to obtain a comprehensive understanding on the effectiveness of a mutant reduction strategy. This may lead to a misjudging on the actual effectiveness of mutant reduction strategies and hence may affect their use in software development.

In order to tackle the above problem, in this study, we aim to propose evaluation indicators to measure to what extent a mutation reduction strategy can maintain the "relative" order of the test effectiveness among test suites. More specifically, for a test suite on a SUT with a set of original mutants, we leverage the test suite to generate a group of test suites that have a partial order relationship in fault detecting potential. When evaluating a reduction strategy, we first construct two partial order relationships among the generated test suites in terms of mutation score, one with the original mutants and another with the reduced mutants. Then, we measure to what extent the two partial order relationships are consistent. The more consistent the two partial order relationships are, the stronger the Order Preservation (*OP*) of the mutation reduction strategy is, and the more effective the reduction strategy is. Furthermore, we propose Effort-aware Relative Order Preservation (EROP) to measure how much gain a mutation reduction strategy can provide compared with a random reduction strategy. Based on a number of open-source programs, we investigate the usefulness of OP and EROP in understanding the effectiveness of various mutation reduction strategies. The experimental results show that OP and EROP can complement existing evaluation indicators for distinguishing various mutation reduction strategies. In addition, we find that Subsuming Mutant Selection (SMS) and Clustering Mutant Selection (CMS) are more effective than the other strategies under OP and EROP.

In this study, we make the following contributions:

- We conduct an in-depth analysis on the pitfalls of existing indicators for evaluating the effectiveness of mutant reduction strategies. We show that, for a mutant reduction strategy, the existing evaluation indicators indeed measure its ability to maintain the "absolute" effectiveness, indicated by the mutation score under the original mutants, of a single test suite. However, they are unable to depict its ability to maintain the "relative" test effectiveness order among test suites. This leads to an incomprehensive (or even counter-intuitive) understanding on the actual effectiveness of a mutant reduction strategy.

- We propose two indicators to evaluate the effectiveness of a mutant reduction strategy: OP and EROP. Different from the existing evaluation indicators, OP and EROP are able to measure the ability of a mutant reduction strategy to maintain the "relative" order of the test effectiveness among different test suites. To the best of our knowledge, this is the first time to take into account measuring the "order-preserving ability" of a reduction strategy. By revealing the pitfalls of the existing indicators, both experimentally and analytically, we seriously suggest that OP and EROP should be taken as the main evaluation indicators in the review of previous studies and the exploration of future work on mutant reduction.

- We conduct an experiment to investigate the effectiveness of five mainstream mutation reduction strategies with the existing and our proposed indicators. The investigated strategies include RMS (Random Mutant Selection), COS (Certain Operator Selection), SMS (Subsuming Mutant Selection), CMS (Clustering Mutant

Selection), and the state-of-the-art mutation reduction strategy Sentinel. The experimental results show that OP and EROP are better than the existing evaluation indicators for distinguishing various mutation reduction strategies. In particular, SMS and CMS are found to be more effective than the other strategies under OP and EROP.

The rest of this paper is organized as follows. Since our main goal is to propose new reduction strategy evaluation indicators, in Section 2, we introduce the relevant background, including mutation testing, mutant reduction, and existing evaluation indicators. In Section 3, we analyze the pitfalls of the existing evaluation indicators and propose our solutions. Section 4 presents the data sets used in this study and the experimental design. Section 5 reports the experimental results. Section 6 discusses the experimental results. Section 7 presents the implications of our study. Section 8 analyzes the threats to the validity of our study. Section 9 concludes the paper and outlines directions for future work.

## 2 BACKGROUND

In this section, we first introduce the mutation testing. Then, we introduce the mutant reduction. Finally, we overview the common indicators for evaluating the effectiveness of a mutant reduction strategy.

### 2.1 Mutation testing

Mutation testing is a testing technology that mutates (changes) statements in SUT to obtain faulty versions (i.e. mutants) to simulate real faults. The mutants are generated automatically by a series of pre-defined mutation operators (i.e. mutators). A test case is said to kill a mutant when the test result differs between a mutant and the SUT according to the strong mutation testing criterion. A mutant is said to be alive (i.e. survived) if and only if none of the test case kills it. There are two concepts of mutation testing: strong mutation testing and weak mutation testing. In the first concept, killing means that the mutant and the original program generate different outputs. While in the second concept, killing means that the program state of the mutant differs from the original one at some point during execution. When it refers to mutation testing, we mean strong mutation testing in this paper. Mutation score is a commonly used indicator to measure the effectiveness of a test suite, which is computed by dividing the number of killed mutants by the total number of non-equivalent mutants:

$$MS(M,T) = \frac{|KM(M,T)|}{|M|} \tag{1}$$

where $M$ is a set of mutants and $T$ is a test suite (i.e. a set of test cases). $MS(M, T)$ is the mutation score computed when executing $T$ against $M$, $KM(M, T)$ is the set of mutants in $M$ killed by $T$, and $|M|$ is the number of mutants in $M$. A high mutation score means that the existing test suite can kill most of the mutants, which makes us believe that it can detect real faults well [3, 4, 12-15]. Equivalent mutants refer to those mutants which are semantically equivalent to the original program. Sometimes, researchers exclude the equivalent mutants, which will never be killed by any test case. However, it is undecidable to determine whether a mutant is equivalent. To this concern, we do not filter the equivalent mutants in this paper.

With the support of mutation score, developers can quantify the "absolute" effectiveness of a single test suite. This enables to determine whether a test suite has a weak fault detecting potential. If it is, there is a need to develop new test cases that are able to kill "alive" mutants in order to enhance its test effectiveness [26]. For a group of test suites, developers can leverage their mutation scores to obtain a "relative" test effectiveness order. If these test suites are generated by different test-generation techniques, this "relative" order will provide a means to determine which test-generation techniques are superior [21]. As can be seen, for practitioners, both the "absolute" effectiveness score and the "relative" effectiveness order are important for understanding the test effectiveness of a test suite. For test suites, once a set of mutations is given, their "absolute" effectiveness scores and "relative" effectiveness order will hence be determined.

**2.2 Mutant reduction**

Considering that the number of mutants is often hundreds to thousands, the overhead of executing all mutants may be unacceptable. Therefore, reducing the cost by mutant reduction has become a research hotspot, which can be defined as to use a mutation reduction strategy $S$ to find a subset $M_s$ of all mutants $M$. Currently, there are two main practices for reduction strategies: one is leveraging the characteristics of the mutants to select the mutants with an execution value, while the other is selecting a mutant subset to maximum a pre-defined optimization function.

For the first practice, the core idea is to leverage the characteristics of mutants that are worth preserving to reduce mutants. The rationale is that some mutants are more relevant to faults or some mutants are more representative. From this perspective, several reduction strategies have been proposed:

- COS (Certain Operator Selection) [3-7, 23]: select the mutants generated by a specific mutator subset of all mutators. There are numerous practices on COS whose core idea is to remove redundant mutation operators and select important operators that are more related to faults. Mathur first proposed COS in 1991 and call it selective mutation [23]. They found that the number of redundant mutants can be reduced by 30% to 40% by deleting "arithmetic operator replacement" mutator and "scalar variable replacement" mutator. Offutt et al. found five important mutation operators for COS, including the relational, logical, arithmetic, absolute, and unary insertion operators [4].

- SMS (Subsuming Mutant Selection) [10-13]: select the subsuming mutants. By the definition in [13], "One mutant subsumes another if at least one test kills the first and every test that kills the first also kills the second". For all mutants, killing the subsuming mutants is to kill all the killable mutants. Therefore, the subsumed mutants should be regarded as redundancy. The practical difficulty of SMS lies in how to figure out the "true" subsumption relationships among all the mutants. In [13], Kurtz et al. proposed both dynamic and static methods to approximate the "true" subsumption. In [11], Gong et al. manually inserted mutant branches into the original program to generate a giant program which is used to analyze for the approximation of the "true" subsumption.

- CMS (Clustering Mutant Selection) [24]: select one mutant from each mutant cluster. First, for all the mutants, a number of features are collected. Second, the data for the mutant instances is used as the input of a clustering algorithm to implement a classification model. Third, by the output of the clustering algorithm, the mutants can be grouped into different clusters. Fourth, the selection is generated by selecting one mutant from each mutant cluster. For example, Hussain [24] executed all the mutants to obtain the information on if a test can kill a mutant or not against all the tests and mutants. After that, K-means and Agglomerative clustering were applied to cluster the mutants. By the clustering algorithm, the mutants in the same cluster were guaranteed to be killed by a similar set of test cases. As a result, randomly selecting one mutant from each cluster can be seen as a selection of representative mutants.

For the second practice, the following approximation is one of the most used optimization functions:

$$MS(M_s, T) \approx MS(M, T) \tag{2}$$

The meaning of this objective function is clear: for a given test suite *T*, the mutation score on the reduced mutant set should be as close as possible to the mutation score on the original mutant set. By (2), the second practice usually attempts to search the mutant subset with an approximate mutation score to the original mutation score, against the given test suite. The rationale behind setting such an objective function is that the criterion of mutant reduction is to keep the evaluation of a test set consistent, which is approximated by mutation score. Considering the wide use of this idea, there are several mutant reduction strategies discussed can be classified in this practice:

- RMS (Random Mutant Selection) [9, 16-18]: select a specified number or proportion of mutants from all mutants randomly. For such a simple strategy, the selecting process of RMS does not need the guidance of an objective function. However, it is a truth that RMS can maintain the mutation score in mathematical expectation (please see Section 3.2.1 for detail). As a result, we can regard the objective function (2) is implied by RMS.
- ROS (Random Operator Selection) [19]: randomly select a specified number or proportion of operator to generate mutants. Similar to RMS, the objective function (2) is implied by ROS, as ROS also can maintain the mutation score in mathematical expectation.

*Strategy effectiveness* is another objective for the pre-defined optimization function, which is only used by the state-of-the-art mutation reduction strategy Sentinel [8]. Considering the complex definition of *strategy effectiveness*, we will elaborate it in the next subsection. Here, we give a brief introduction to Sentinel:

- Sentinel [8]: generate the cost-optimal reduction strategies instead of selecting mutants directly. To search the possible strategy space, Sentinel defines two objective functions. One function is to minimize the execution time. The other function is to maximize the *strategy effectiveness*. With several pre-defined basic mutant selection (i.e. mutant reduction) strategies, Sentinel searches a combination of basic strategies instead of selecting mutants directly for the training project which can achieve the best values of the objective functions.

After a brief introduction to several major reduction strategies, we will describe how the existing work evaluated these strategies in the next section.

**2.3 Mutant reduction evaluation indicator**

In evaluating a reduction strategy, researchers are most concerned with the following two questions: how many mutants have been reduced? does the evaluation on the given test suite change significantly before and after reduction? As a result, they mainly evaluate a reduction strategy from two dimensions: one is the number of reductions, while the other is the effect of reductions on test suite effectiveness evaluation.

To hit the first dimension, the simplest indicator is the Reduction Ratio *RR* of mutants:

$$RR = \frac{|M|-|M_s|}{|M|} \quad (3)$$

Besides, instead of directly evaluating the number of reduced mutants, a computation on the reduction of executing time can be used as an alternative [8]. The meaning of these indicators is clear: the more reduction, the less cost. In the literature, a tiny minority of studies are evaluated only by using reduction ratio [22]. Considering that the meaning of the indicators for the first dimension is simple and clear, this paper will focus on the indicators for the second dimension.

To hit the second dimension, the variation of mutation score (i.e. *VMS = |MS(M, T)- MS($M_s$, T)|*) is the most commonly used indicator in the majority of existing research [4,6-11]. For the existing studies, the smaller the variation, the better the effect is considered. It is worth noting that many existing studies emphasized that the projects they studied had a perfect test suite (i.e. *MS (M, T)* is close to 100%) [4, 6-11]. As a result, it is also an alternative

to directly use the mutation score after reduction as an indicator. The higher mutation score after reduction, the better the effect is. The meaning of this indicator seems to be justified: the smaller the variation, the smaller the change in the "absolute" evaluation on the given test suite. If a reduction strategy can provide a small variation on most experimental projects, in practice, we have a reason to believe that for any test suite, the "absolute" evaluation deviation caused by the reduction is small.

However, the use of variations in mutation score had also been questioned. Gopinath pointed out that "operator selection, and stratified sampling (on operators or program elements) — are at best marginally better than random sampling and are often worse" by using the *strategy effectiveness* ($E_s$) [20]. Assume that $T$ is the test suite on the SUT with the original mutant set $M$. For a target mutant reduction strategy $S$, let $M_s$ be the set of the mutants reduced from $M$. From the original test suite $T$, we select all the test cases that can detect any mutants in $M_s$ and finally get a set $T_s$. Then, $E_s$ is defined as follows:

$$E_s = \frac{|KM(M,T_s)|}{|KM(M,T)|} - E_r \tag{4}$$

Here, $KM(M, T_s)$ is the set of mutants in $M$ killed by $T_s$, $KM(M, T)$ is the set of mutants in $M$ killed by $T$. In particular, $E_r$ is the effectiveness of a reduction strategy that randomly selects a set $M_r$ of mutants from M with the condition $|M_r| = |M_s|$, i.e.,

$$E_r = \frac{|KM(M,T_r)|}{|KM(M,T)|} \tag{5}$$

where $T_r$ is the set of test cases in $T$ that can kill any mutant in $M_r$ and $KM(M, T_r)$ is the set of mutants in $M$ killed by $T_r$. In nature, $E_s$ depicts the mutant killing ability of the test set $T_s$ associated with a reduction strategy relative to a random selection strategy that selects the same number of mutants. The higher the mutant killing ability of the test set $T_s$ is, the higher $E_s$ a reduction strategy has. In the following, for the simplicity of presentation, we call $E_s$ in (4) as the "relative" strategy effectiveness. In particular, we call $E_s$ without the second item $E_r$ as the "absolute" strategy effectiveness. In default, $E_s$ denotes the "relative" strategy effectiveness.

As can be seen, *VMS* and $E_s$ do not explicitly measure the degree to which a mutation reduction strategy can maintain the relative order of the effectiveness among test suites. That is to say, they may answer the question: does the "absolute" evaluation on a given test suite change significantly before and after reduction? However, they cannot answer the question: does the "relative" effectiveness order among test suites change significantly before and after reduction? As we mentioned before, "by mutation score, developers can not only know the 'absolute' effectiveness of a given test suite, but also know the 'relative' order of the effectiveness among different test suites." In the next section, we will figure out the pitfalls of ignoring the "relative" effectiveness measurement and explore the feasible solutions.

## 3 MUTANT REDUCTION EVALUATION: ESSENTIAL, PITFALLS, AND SOLUTIONS

In this section, we first analyze the essential of mutation reduction evaluation from the viewpoint of measurement theory. Then, we point out the pitfalls of the existing reduction evaluation indicators. Finally, we propose our solutions to mutation reduction evaluation.

## 3.1 Essential of mutation reduction evaluation

For the simplicity of presentation, we make the following assumptions for the mutation reduction evaluation problem: (1) $M$ is the original mutation set; (2) $S$ is a mutant reduction strategy; and (3) $M_s$ is the resulting reduced mutant set by $S$. In this context, there is a need to use an indicator *REI* (Reduction Evaluation Indicator) to evaluate the reduction effectiveness of $S$. With such an indicator *REI*, practitioners could answer the following important question: how well is the reduced mutant set $M_s$ in maintaining the ability of test suite effectiveness evaluation of the original mutant set $M$?

Under the given original mutant set $M$, for each test suite $T$, we can compute the corresponding mutation score $MS(M, T)$. From the viewpoint of measurement theory, mutation score $MS$ is a mapping (i.e. metric) $MS(M, .): E \to V$ which yields for each test suite $T \in E$ a measurement value $MS(M, T) \in V$. If $MS(M, .)$ is a valid metric, this means that there is a scale $<ERS, NRS, MS>$. Here, $ERS = (E, \preceq)$ is an empirical relationship system, where $E$ is a set of test suites and $\preceq$ is a binary empirical relationship on $E$. $NRS = (V, \leq)$ is a numerical relationship system, where $V$ is the set of real numbers and $\leq$ is a binary numerical relationship on $V$. In particular, $MS(M, .)$ satisfies the following representation condition: $T1 \preceq T2 \Leftrightarrow MS(M, T1) \leq MS(M, T2)$ for all $T1, T2 \in E$. In other words, the binary empirical relation $\preceq$ is preserved in $V$ by $MS(M, .)$.

What does mutation reduction mean from the viewpoint of measurement theory? Assume that $MS(M, .)$ is a valid metric and the binary relationship $\preceq$ denotes "less effective than or as effective as" with respect to test effectiveness. Furthermore, assume that the mutation reduction strategy $S$ selects a subset $M_s$ from the mutation set $M$. Consequently, we will obtain a new mutation score metric $MS(M_s, .)$. Recall that the purpose of mutation reduction is to reduce the computation cost of mutation score but maintain the ability of test suite effectiveness evaluation (abbreviated as "reduce computation cost but maintain evaluation ability"). Ideally, we hope that $MS(M_s, .)$ is also a valid metric, i.e., $T1 \preceq T2 \Leftrightarrow MS(M_s, T1) \leq MS(M_s, T2)$ holds for all $T1, T2 \in E$. In other words, the binary empirical relation $\preceq$ is also preserved in $V$ by $MS(M_s, .)$. In this case, $MS(M, .)$ and $MS(M_s, .)$ produce the same numerical relationship on $V$. This means that, compared with $MS(M, .)$, $MS(M_s, .)$ causes a zero-loss in the ability of test suite effectiveness evaluation.

However, in practice, it is very possible that, after mutation reduction, a loss will be caused in the ability of test suite effectiveness evaluation. In this case, how to evaluate the effectiveness of a mutation reduction strategy S? A natural idea is to compare the numerical relationships on $V$ produced by $MS(M, .)$ and $MS(M_s, .)$. If there is a large difference, this means that the numerical relationship produced by $MS(M, .)$ is largely twisted by mutation reduction, i.e., there is a large change in the ability to evaluate test suite effectiveness. If there is a small difference, this means that $MS(M, .)$ and $MS(M_s, .)$ have a similar ability to evaluate test suite effectiveness. Given this situation, we can use the difference between the numerical relationships on $V$ before and after reduction to measure the effectiveness of the mutation reduction strategy $S$: the smaller the difference is, the more effective the reduction strategy is. Since the numerical relationship in our context is a binary relationship, the essential of mutation reduction evaluation is to measure the "order-preserving ability", i.e., to what extent the mutation score order among test suites is maintained before and after mutation reduction.

## 3.2 Pitfalls of existing evaluation indicators

According to Section 3.1, the essential of mutation reduction evaluation is to measure the extent of change in mutation score order among test suites before and after mutation reduction. During this measurement, the mutation score order among test suites before mutation reduction is regarded as the ground truth. However, the existing REIs

| kill($m_i$, $t_j$) | test $t_1$ | test $t_2$ | test $t_3$ | test $t_4$ |
|---|---|---|---|---|
| mutant $m_1$ | 1 | 0 | 0 | 0 |
| mutant $m_2$ | 0 | 0 | 1 | 0 |
| mutant $m_3$ | 1 | 1 | 1 | 0 |
| mutant $m_4$ | 1 | 1 | 1 | 1 |
| mutant $m_5$ | 0 | 0 | 0 | 1 |

Selected mutants given by Strategy A: $m_1$, $m_2$
Selected mutants given by Strategy B: $m_3$, $m_4$

$$\text{kill}(m_i, t_j) = \begin{cases} 1, & \text{if } t_j \text{ kills } m_i \\ 0, & \text{else} \end{cases}$$

Figure 1. An example to show the misleading of strategy effectiveness.

such as *VMS* and $E_s$ do not directly measure such a change in mutation score order among test suites. In contrast, for a mutation reduction strategy *S*, they quantify either its influence on the mutation score of the same test suite (*VMS*), or the ability of the tests killing the reduced mutants to kill the original mutants ($E_s$). Clearly, if S has a high *VMS* and/or $E_s$, this does not necessarily imply that it also has a high order-preserving ability. In other words, it is unable to use *VMS* and $E_s$ to determine how close are the mutation score orders among test suites before and after reduction. Consequently, misleading conclusions could be possibly achieved when using the existing REIs to evaluate the reduction effectiveness. This is especially true when comparing the effectiveness of multiple reduction strategies.

We next use an example to illustrate the pitfalls of *VMS* and $E_s$. Fig. 1 shows a kill relationship matrix for 5 mutants and 4 test cases. The matrix reflects the kill relationship between test cases and mutants: kill(*m*, *t*)= 1 if and only if *t* kills *m*. In this example, the original mutation set $M = \{m_1, m_2, m_3, m_4, m_5\}$, the original test suite $T = \{t_1, t_2, t_3, t_4\}$. Assume that there are two mutation reduction strategies A and B: according to A, the reduced mutant set $M_A = \{m_1, m_2\}$; according to B, the reduced mutant set $M_B = \{m_3, m_4\}$. Now, the problem we need to answer is: is A or B more effective in mutation reduction? As can be seen, *MS*(*M*, .), *MS*($M_A$, .), and *MS*($M_B$, .) are indeed three metrics corresponding to the original mutation set *M*, the reduced mutant set $M_A$, and the reduced mutant set $M_B$, respectively. According to Section 3.1, if a reduction strategy leads to a metric that has a high "order-preserving ability", it will be regarded as a good reduction strategy. Therefore, in order to compare the reduction effectiveness of A and B, we need to analyze the "order-preserving ability" of *MS*($M_A$, .) and *MS*($M_B$, .) compared with *MS*(*M*, .). Before this analysis, there is a need to know the mutation score order among test suites by *MS*(*M*, .). To this end, we leverage the original test suite T to generate all possible non-empty subsets, i.e., $2^4-1=15$ non-empty subsets. Fig. 2(a) describes the mutation score order under *MS*(*M*, .) among the test suites that have a subset relationship[1]. For each arrow, the test suite on the head is a subset of the test suite on the tail. In particular, a green arrow indicates that the test suite on the tail has a higher mutation score, while a yellow arrow indicates that these two test suites have the same mutation score. Fig. 2(b) and Fig. 2(c) respectively describe the mutation score order under *MS*($M_A$, .) and *MS*($M_B$, .) among the test suites that have a subset relationship. For a pair of test suites with a subset relationship, if the mutation score order is changed compared with that under *MS*(*M*, .), the corresponding arrow is shown in red. When comparing Fig. 2(b) with Fig. 2(a), we can find that 7 out of 28 arrows are red. When comparing Fig. 2(c) with Fig. 2(a), we can find that 19 out of 28 arrows are red. Therefore, from the viewpoint of "order-preserving ability", both A and B are not perfect, but A is a more effective reduction strategy than B. In the following, we will use this example to investigate whether *VMS* and $E_s$ could achieve the same conclusion.

---

[1] It is worth noting that we limit our analysis to the test suites having a subset relationship, as this can ensure that the mutation score order among test suites under *MS*(M, .) can be regarded as the ground truth.

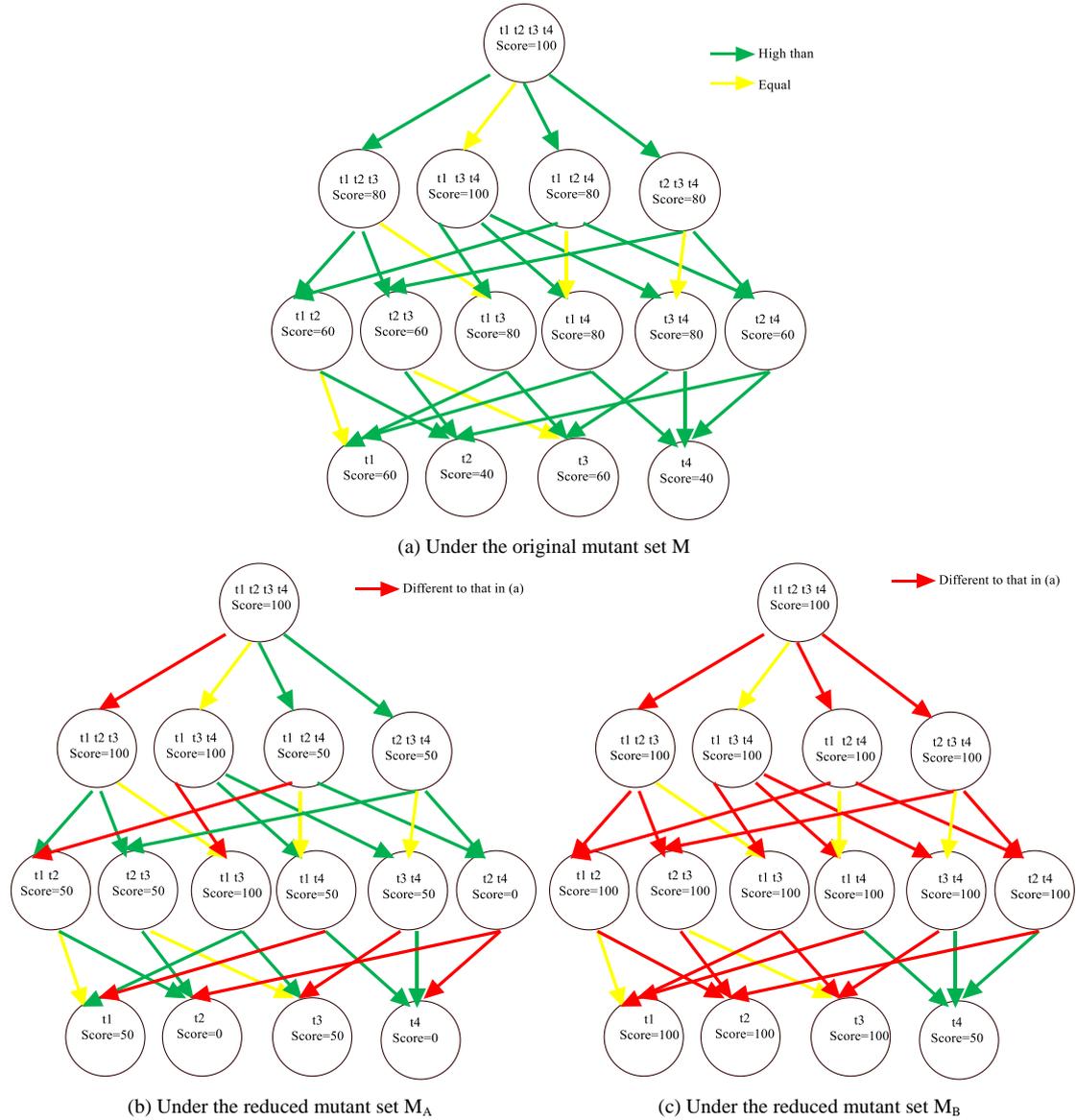

Figure 2. The visualization of mutation score order among test suites (generated from the test suite shown in Fig. 1) under different mutant sets

### 3.2.1 Pitfalls of variation of mutation score

From Fig. 1, we can see that: $VMS(A) = |MS(M, T) - MS(M_A, T)| = |1 - 1| = 0$ and $VMS(B) = |MS(M, T) - MS(M_B, T)| = |1 - 1| = 0$. Therefore, according to *VMS*, both A and B are perfect in maintaining the ability of test suite effectiveness evaluation and A is as effective as B. Compared with the conclusion drawn from the "order-preserving ability", we can find that: on the one hand, *VMS* is unable to reveal the fact that both $MS(M_A, .)$ and $MS(M_B, .)$ twist the mutation score order among test suites under $MS(M, .)$; on the other hand, *VMS* is unable to reveal the fact that $MS(M_A, .)$ has a higher "order-preserving ability".

***Why does the variation of mutation score go wrong?*** Indeed, a mutation reduction strategy that can keep the mutation score unchanged is of course the best. However, it is worth noting that it must be maintained for ALL test suites. That is to say, when using mutation score to guide mutant reduction, instead of using (2), we should use:

$$\text{For ANY } T, \; MS(M_s, T) = MS(M, T) \tag{5}$$

For a reduction strategy, if (5) holds, it is natural that the mutation score order under *M* is completely preserved under $M_s$, i.e., $M_s$ has the same ability to evaluate test suite effectiveness as *M*. Obviously, such a requirement is strict, which means that it is hard to satisfy (5) in practice. As a result, many existing studies use *VMS* to characterize the difference in mutation score before and after reduction. In particular, it is a common practice to compute *VMS* based on a number of test suites having a high *MS(M, .)* (such as around 100%, sometimes called mutation-adequate test suites [4, 6-11]) and believe that a small *VMS* indicates a better a reduction strategy. However, there are three pitfalls in this practice. First, a small *VMS* on mutation-adequate test suites cannot assure that a small *VMS* on non-mutation-adequate test suites, which are very common for real-world programs [29]. In other words, a good reduction strategy selected by *VMS* on mutation-adequate test suites may not be applicable to non-mutation-adequate test suites. Second, *VMS* on mutation-adequate test suites may have a low ability to distinguish between different reduction strategies. As an example, if *MS(M, T)* = 100%, this means that all the mutants in *M* can be killed by *T*. In this context, any mutant reduction strategy will lead to $MS(M_s, T)$ = 100%. Consequently, *VMS* is unable to distinguish between any two reduction strategies. This is a possible reason why many prior studies report that many reduction strategies have a similar reduction effectiveness [29]. Third, most importantly, for a reduction strategy S, even if its *VMS* is small for all test suites, it does not necessarily mean that $MS(M_S, .)$ has a good "order-preserving ability". For example, for two test suites *T*1 and *T*2, assume that *MS(M, T*1*)* = 90%, *MS(M, T*2*)* = 89%, $MS(M_s, T1)$ = 89%, and $MS(M_s, T2)$ = 90%. As can be seen, although *VMS*s on *T*1 and *T*2 are both trivial, the mutation score order on *T*1 and *T*2 is inverse before and after mutation reduction. The fundamental reason for the above pitfalls, we believe, is that *VMS* aims to measure the "mutation-score-preserving ability" of a reduction strategy, which is not coincident exactly with the "order-preserving ability".

Another noteworthy fact is that a random reduction strategy RMS always maintains the following property:

$$E[MS(M', T)] = MS(M, T) \tag{6}$$

where *M'* is the set of mutants randomly selected from *M* by RMS and *E* is a mathematical expectation. In the following, we give a brief proof. Assume that $|M| = n$, $|M'| = m$, and $|KM(M, T)| = k$. Then, we have:

$$E[MS(M', T)] = E\left[\frac{|KM(M', T)|}{|M'|}\right]$$

$$= \frac{E[|KM(M', T)|]}{|M'|}$$

$$= \frac{1}{m} E[|KM(M', T)|]$$

$$= \frac{1}{m} \frac{1}{C_n^m} \sum_{i=0}^{k} i \, C_k^i C_{n-k}^{m-i}$$

$$= \frac{1}{m} \frac{1}{C_n^m} \sum_{i=1}^{k} i \, C_k^i C_{n-k}^{m-i}$$

$$= \frac{1}{m} \frac{1}{C_n^m} \sum_{i=1}^{k} i \frac{k!}{(k-i)! \, i!} C_{n-k}^{m-i}$$

$$= \frac{1}{m} \frac{1}{C_n^m} \sum_{i=1}^{k} k \frac{(k-1)!}{(i-1)! \, ((k-1)-(i-1))!} C_{n-k}^{m-i}$$

$$= \frac{1}{m}\frac{1}{C_n^m}k\sum_{i=1}^{k} C_{k-1}^{i-1} C_{n-k}^{m-i}$$

$$= \frac{1}{m}\frac{C_{n-1}^{m-1}}{C_n^m}k$$

$$= \frac{1}{m}\frac{m}{n}k$$

$$= \frac{k}{n}$$

$$= MS(M,T)$$

This means that RMS always keeps the mutation score unchanged in expectation. Under *VMS*, it is natural that all pretty strategies are similar to a random strategy. However, from the viewpoint of RMS, all mutants have the same weight. That is to say, RMS focuses on the number of reductions more than the effect of reductions. If an indicator on the effect of reductions gives a high evaluation for the strategy that does not pay much attention to the effect of reductions, the rationality of this indicator maybe questionable.

*3.2.2 Pitfalls of strategy effectiveness*

When applying the strategy effectiveness $E_s$ to the reduction strategies A and B, we have:

$$E_s(A) = \frac{|KM(M,T_s(A))|}{|KM(M,T)|} - E_r, \ E_s(B) = \frac{|KM(M,T_s(B))|}{|KM(M,T)|} - E_r$$

Because of the same *KM (M, T)* and $E_r$ (the strategy effectiveness of a random model), we can find that the part that really affects the $E_s$ is $|KM(M, T_s)|$. From the definition of *KM (M, T_s)*, we can see that the larger $|T_s|$, the $|KM|$ is more likely to be large. In other words, the more selected test cases (i.e. test cases that can kill any selected mutants), the higher $|KM(M, T_s)|$. and $E_s$. From Fig. 1, we can see that: $T_s(A) = \{t_1, t_3\}$ and $T_s(B) = \{t_1, t_2, t_3, t_4\}$. Since $T_s(A)$ is a subset of $T_s(B)$, we have:

$$|KM(M,T_s(A))| = |\{m_1,m_2,m_3,m_4\}| \ < \ |KM(M,T_s(B))| = |\{m_1,m_2,m_3,m_4,m_5\}|.$$

Therefore, we can conclude that $E_s(A) < E_s(B)$. In particular, we can observe that $|KM(M,T_s(B))| = |KM(M,T)|$, indicating that $E_s(B)$ achieves the maximum strategy effectiveness. Therefore, according to $E_s$, B is perfect in maintaining the ability of test suite effectiveness evaluation and A is less effective than B. Compared with the conclusion drawn from the "order-preserving ability", we can find that: on the one hand, $E_s$ is unable to reveal the fact that $MS(M_B, .)$ twists the mutation score order among test suites under $MS(M, .)$; on the other hand, $E_s$ is unable to reveal the fact that $MS(M_A, .)$ has a higher "order-preserving ability".

**Why does the Strategy effectiveness make mistake?** In nature, $E_s$ depicts the mutant killing ability of the test set $T_s$. Under the same reduction ratio, the larger $|KM(M, T_s)|$ is, the higher (i.e. better) $E_s$ is. The main pitfall is that the aim of $E_s$ is not directly related to the "order-preserving ability". In other words, a higher $E_s$ does not necessarily mean a higher "order-preserving ability". For example, supposing that there is a mutant m killed by all test cases in the original test suite *T* (e.g. a mutant leading to "time out error") and a reduction strategy *S* selects only m. Then, we have $T_s = T$. Consequently, $KM(M, T_s) = KM(M, T)$. In this case, *S* achieves the maximum strategy effectiveness under the least reduction ratio (i.e. only one mutant is selected). However, by this reduction strategy *S*, for any test suite *T*1 including at least one test case in *T*, we have $MS(\{m\}, T1) = 100\%$. This means that $MS(\{m\}, .)$ is unable to distinguish the test effectiveness for all these test suites. In this sense, the reduction strategy *S* identified by $E_s$ will lose its practical value.

### 3.3 Proposed evaluation indicators: OP and EROP

As shown in Section 3.2, when evaluating the effectiveness of a reduction strategy, the existing mutation reduction evaluation indicators are unable to measure the "order-preserving ability" after reduction. As a result, they may lead to misleading conclusions. In order to tackle this problem, we next propose indicators to quantify the "order-preserving ability" for a mutation reduction strategy $S$. Assume that, by $S$, the original mutation set $M$ on the SUT is reduced to a mutation set $M_s$. As a result, we have two mutation score metrics, $MS(M, .)$ and $MS(M_s, .)$. At a high level, we first leverage the original test suite T to generate a group of test suites that have a partial order relationship defined by the subset relation[2]. Then, we generate two mutation score orders between these test suites that have a subset relationship: one from $MS(M, .)$, and another from $MS(M_s, .)$. Finally, we compare them to calculate to what extent the mutation score order produced by $MS(M, .)$ is changed with respect to the mutation score order produced by $MS(M_s, .)$. During this process, the mutation score order produced by $MS(M, .)$ is used as the ground truth. By this way, we know to what extent the ability to evaluate test suite effectiveness is changed after mutation reduction.

During the above process, the computation complexity is mainly from the comparison of two mutation score orders produced by $MS(M, .)$ and $MS(M_s, .)$. In our context, a mutation score order is indeed a set of ($T1$, $T2$, operator) satisfying the following conditions: (1) $T2$ is a proper subset of $T1$; and (2) operator is ">" if $T1$ has a higher mutation score and is "=" if $T1$ has the same mutation score compared with $T2$. For each pair of test suites, we examine whether the operator on ($T1$, $T2$, operator) is changed before and after reduction. Assume that $T1$ is a subset of $T2$ and $T2$ is a subset of $T3$. Clearly, it is redundant to compare $T1$ with $T3$ if we have compared $T1$ with $T2$ and $T2$ with $T3$. In our study, in order to simplify the computation, we only take into account the following mutation score comparisons: $T1$ vs. $T2$ and $T2$ vs. $T3$. With the above constraint, how many pairs of test suites we need to examine at least for comparing the two mutation score orders produced by $MS(M, .)$ and $MS(M_s, .)$? Let P(n) be the minimum total number of pairs of test suites we need to examine, where n is the number of test cases in the original test suite $T$. Suppose that SNES($T$) is the Set of Non-Empty Subsets of $T$ (each element is a set of test cases, i.e., a test suite). As a result, we have |SNES($T$)| = $2^n$-1. For the simplicity of presentation, if $T1$ is an element of SNES($T$) and $T1$ consists of k test cases, then we call $T1$ a k-subset of $T$ ($1 \leq k \leq n$). For each k-subset T1 with k > 1, there are k k-1-subsets, each one corresponding to one set by removing one single test case from $T1$. In this context, in order to compute P(n), for each k-subset with k > 1, we need to compare it against each of the corresponding k k-1-subset, i.e., there are k comparisons involved in total. Since there are $C_n^k$ k-subsets, therefore, we have:

$$P(n) = \sum_{k=2}^{n} k \times C_n^k \tag{7}$$

Through simplification, we can get:

$$P(n) = \sum_{k=2}^{n} k \times C_n^k = \sum_{k=2}^{n} n \times C_{n-1}^{k-1} = n \times \sum_{k=2}^{n} C_{n-1}^{k-1} = n \times (2^{n-1} - 1) \tag{8}$$

---

[2] In the literature, for a SUT, it is often the case that only one test suite (i.e., the original test suite T) is provided. We use T to generate a group of test suites with a partial order relationship on test effectiveness. This ensures, on the one hand, our proposed evaluation indicators can be easily applied in practice, and on the other hand, the mutation score order produced by $MS$(M, .) reflects the Score Monotonicity Property: adding test cases to a test suite does not decrease its mutation score [25], which can be used as the ground truth.

In the following, for the simplicity of presentation, such a calculation approach is called a "continuous subsample" approach. Taking Fig. 2(a) as an example, there are 4 3-subsets, 6 2-subsets, and 4 1-subsets. As a result, 1*4+2*6+3*4=28 comparisons (i.e. arrows) are needed, which can be computed by $P(4) = 4 \times (2^3 - 1)$.

As shown in equation (8), the computation of P(n) has an exponential time complexity, which hinders the quantification of the "order-preserving ability" for a mutation strategy in practice. For the sake of practicability, we need to reduce the number of comparisons. To this concern, for comparing two mutation score orders produced by $MS(M, .)$ and $MS(M_s, .)$, in this study, we hence propose to use the following "continuous half-sample" approach to generate a group of test suites that have a partial order relationship defined by the subset relation:

(1) $T_0 = T$, i = 0;
(2) If $|T_i| = 1$, then stop, else go to (3)
(3) Generate $T_{i+1}$ by randomly selecting half of the test cases (round down) in $T_i$;
(4) i = i + 1;
(5) Go to (2).

Consequently, we obtain the following k test suites:

$$T_0 \supset T_1 \supset T_2 \supset \cdots \supset T_k \neq \emptyset$$

which satisfies:

$$|T_{i+1}| = int(|T_i| \times 0.5), i = 0, 1, ..., k-1$$

and

$$k = int(log_2|T_0|) = int(log_2 n)$$

For such a sequence of test suites, we only need to compare k times, far less than $n \times (2^{n-1} - 1)$, to determine the change in two mutation score orders produced by $MS(M, .)$ and $MS(M_s, .)$. After computing $MS(M, T_i)$ and $MS(M_s, T_i)$ for all i, we have:

$$MS(M, T_{i+1}) < MS(M, T_i) \text{ or } MS(M, T_{i+1}) = MS(M, T_i) \ (i=0, 1, ..., k-1)$$

and

$$MS(M_s, T_{i+1}) < MS(M_s, T_i) \text{ or } MS(M_s, T_{i+1}) = MS(M_s, T_i) \ (i=0, 1, ..., k-1)$$

Although there are two possible signs, there is only one sign in fact. We record the 2×k signs according to the mutation score order. By the 2×k relationships, we obtain the set of index of changed signs:

$$X = \{i \mid MS(M_s, T_{i+1}) = MS(M_s, T_i) \text{ and } MS(M, T_{i+1}) < MS(M, T_i)\} \cup$$
$$\{i \mid MS(M_s, T_{i+1}) < MS(M_s, T_i) \text{ and } MS(M, T_{i+1}) = MS(M, T_i)\}$$

However, indeed, if $MS(M, T_{i+1})$ and $MS(M, T_i)$ are the same, $MS(M_s, T_{i+1})$ and $MS(M_s, T_i)$ must be the same (i.e. the latter set is an empty set). This is because if $T_{i+1}$ and $T_i$ cannot be distinguished by M, $T_{i+1}$ and $T_i$ must not be distinguished by a subset of M. For example, the yellow arrows in Fig.2(a) are still yellow in Fig.2(b) and Fig.2(c). As a result, we have:

$$X = \{i \mid MS(M_s, T_{i+1}) = MS(M_s, T_i) \text{ and } MS(M, T_{i+1}) < MS(M, T_i)\}$$

As such, we propose to use the Order Preservation (*OP*) to measure the change in two mutation score orders produced by $MS(M, .)$ and $MS(M_s, .)$ as follows:

$$OP(S) = 1 - |X|/k \qquad (9)$$

As can be seen, there is randomness when generating $T_1, ..., $ and $T_k$ from the given original test suite $T$. In order to reduce the influence of randomness, we recommend that the k test suites are generated 100 times and the resulting average OP is used for evaluating the "order-preserving ability". In the following, when we talk about OP, we mean the final OP computed by average. The higher OP value, the higher the "order-preserving ability" of a mutation reduction strategy (based on a given reduction ratio). In particular, for the simplicity of presentation, such a calculation procedure will be called a "continuous half-sample" approach.

We next use the example in Fig. 2 to demonstrate how to compute *OP*. At first, from the original test suite $T = \{t_1, t_2, t_3, t_4\}$, we generate the following sequence of test suites:

$$T0 = \{t_1, t_2, t_3, t_4\}, T1 = \{t_1, t_2\}, T2 = \{t_1\}.$$

Table 1 reports for each test suite in this sequent the mutation score before and after mutation reduction:

Table 1 Mutation scores under different mutation sets

| MS ($M_i$, $T_j$) | M = {$m_1$, $m_2$, $m_3$, $m_4$, $m_5$} | MA = {$m_1$, $m_2$} | MB = {$m_3$, $m_4$} |
|---|---|---|---|
| T0 | 100% | 100% | 100% |
| T1 | 60% | 50% | 100% |
| T2 | 60% | 50% | 100% |

As can be seen, the mutation score order before reduction is:

$$\{(T0, T1, ">"), (T1, T2, "=")\}$$

The mutation score order after applying the mutation reduction strategy A is:

$$\{(T0, T1, ">"), (T1, T2, "=")\}$$

The mutation score order after applying the mutation reduction strategy B is:

$$\{(T0, T1, "="), (T1, T2, "=")\}$$

As such, the sets of index of changed signs for strategy A and strategy B are respectively X(A) = { } and X(B) = {0}. Therefore, *OP*(A) = 1 − 0/2 =1 and *OP*(B) = 1 − 1/2 = 0.5. By using OP, we can distinguish strategy A from strategy B. For this sequence, the conclusion is strategy A is better than strategy B, which is consistent with our analysis in 3.2. If we use variation of mutation score, we have *VMS*(A) and *VMS*(B) = 0, i.e., there is no way to distinguish strategy A from strategy B.

However, OP is a non-effort-aware evaluation indicator, as it does not take into account the number of reduced mutants. Indeed, a strategy that does not delete any mutant will have an OP of 1.0 (i.e., the maximum). To tackle this problem, in this study, we propose an effort-aware evaluation indicator Effort-aware Relative Order Preservation (EROP) to measure the "order-preserving ability" as follows:

$$EROP(S) = RR(S) * ROP(S) \tag{10}$$

Here, RR(*S*) is the mutation reduction ratio, while ROP(*S*) = OP(*S*) − OP(*random*) is the Relative Order Preservation ability of a reduction strategy *S* compared with a random reduction strategy that selects the same number of mutants. The higher EROP value, the better effectiveness of a strategy. If the value is negative, the strategy is useless, as it is not superior to a random reduction strategy. Similar to OP, we also recommend that the k test suites are generated 100 times to compute the average EROP when evaluating the "order-preserving ability" of a reduction strategy *S* in practice.

In summary, in this study, we propose two indicators, OP and EROP, for evaluating the "order-preserving ability" of a mutation reduction strategy. The former is non-effort-aware, while the latter is effort-aware. For a SUT with the original mutant set *M* and the original test suite *T*, OP and EROP can be efficiently computed. In the following sections, we empirically investigate the actual usefulness of OP and EROP, especially when used to compare the effectiveness of different reduction strategies.

# 4 EXPERIMENTAL SETUP

In this section, we describe in detail the experimental design. First, we report the programs and tools used in our study. Then, we introduce the investigated mainstream mutation reduction strategies and their implementations. Finally, we list the research questions under study.

## 4.1 Subject programs and tools

**Programs.** As shown in Table 2, ten Java programs, which are from Apache projects, are selected for the experiments [28]. We selected these 10 programs for study, as they were widely used in previous work [8,11,26,27,29-31]. They range in size from dozens to thousands of lines. In particular, as indicated by mutation score, there is a relatively sufficient set of test cases against each program. Note that we use a single program rather than an entire project for the following reasons. First, the result of mutant reduction on a project is more accidental. For example, it is possible for random mutant selection to select the mutants within a single class without selecting mutants in the other classes. Second, for a given project, the test class is more closely related to the class under testing than other classes. Due to this fact, the testing process on the whole project can be regarded as the combination of the testing process of each class. As a result, it is more valuable to combine the reduction results of each class as the final reduction result of the whole project. By these considerations, using programs as the experimental objects is more practical.

**Mutants.** The tool we use to generate and run the mutants is PIT 1.4.3 [32], which is a widely used tool for Java mutation testing. To generate the mutants, 7 default mutators are used in this paper (see section 4.2.3 for detail). It is worth noting that one of the aims of our experiment is to evaluate various mutation reduction strategies. Therefore, we need to execute all the mutants. After the mutants are executed, not only the mutation score but also the detail of which tests are executed can be obtained by the PIT reports for each program. Table reports the number of executed tests. From Table 2, we can see that the high mutation scores (average around 0.9) indicate that the test suites are effective. In the execution process, whether each test case can kill a mutant is recorded and used to obtain the kill relationship matrix (e.g. Fig. 1). This matrix is used only when evaluating the strategies and all the studied strategies are not implemented on this information.

Table 2. The information of programs

|     | Program | LOCs | Path | Tests | Mutants | MS |
| --- | --- | --- | --- | --- | --- | --- |
| J1  | Crypt | 25 | org.apache.commons.codec.digest | 12 | 11 | 1.000 |
| J2  | StringUtils | 48 | org.apache.commons.codec.binary | 172 | 39 | 0.974 |
| J3  | Md5Crypt | 107 | org.apache.commons.codec.digest | 16 | 62 | 0.887 |
| J4  | DefaultParser | 178 | org.apache.commons.cli | 64 | 109 | 0.929 |
| J5  | Option | 192 | org.apache.commons.cli | 303 | 111 | 0.800 |
| J6  | ArithmeticUtils | 207 | org.apache.commons.math3.util | 307 | 213 | 0.816 |
| J7  | ResizableDouble Array | 217 | org.apache.commons.math3.util | 60 | 149 | 0.743 |
| J8  | UnixCrypt | 311 | org.apache.commons.codec.digest | 8 | 215 | 0.967 |
| J9  | HelpFormatter | 416 | org.apache.commons.cli | 31 | 134 | 0.881 |
| J10 | Dfp | 1702 | org.apache.commons.math3.dfp | 159 | 1161 | 0.954 |

Table 3. The 5- Operators Selection (in red) among 22 mutators

| Mutator | Description |
|---|---:|
| AAR | array reference for array reference replacement |
| ABS | <span style="color:red">absolute value insertion</span> |
| ACR | array reference for constant replacement |
| AOR | <span style="color:red">arithmetic operator replacement</span> |
| ASR | array reference for scalar variable replacement |
| CAR | constant for array reference replacement |
| CNR | comparable array name replacement |
| CRP | constant replacement |
| CSR | constant for scalar variable replacement |
| DER | DO statement end replacement |
| DSA | DATAstatement alterations |
| GLR | GOTO label replacement |
| LCR | <span style="color:red">logical connector replacement</span> |
| ROR | <span style="color:red">relational operator replacement</span> |
| RSR | RETURN statement replacement |
| SAN | statement analysis |
| SAR | scalar variable for array reference replacement |
| SCR | scalar for constant replacement |
| SDL | statement deletion |
| SRC | source constant replacement |
| SVR | scalar variable replacement |
| UOI | <span style="color:red">unary operator insertion</span> |

**Coverage.** The tool we use to collect the coverage information is Cobertura [33], which is a widely used coverage tool. All the tests recorded by PIT reports will be executed against SUT by Cobertura to obtain the coverage relationship matrix, which records whether each test case can cover a mutant or not. In the following, we say a test case covers a mutant if and only if the test case covers the corresponding statement to be mutated in the original program SUT.

**Tests.** For each target program, the manually written test suite provided by Apache projects according to the SUT is the raw test suite. However, not all test cases in the raw test suite are used. In our study, only the test cases which cover the target program and execute successfully by PIT are used.

### 4.2 Five mutation reduction strategies and their implementations

We use Random Mutant Selection (RMS), Sentinel, Certain Operator Selection (COS), Subsuming Mutant Selection (SMS), and Clustering Mutant Selection (CMS) as the mainstream mutation reduction strategies in the experiment. Given a SUT, the corresponding mutants, and a test suite, since the uncovered mutants must be alive, we first use the coverage information to filter out uncovered mutants which do not need to be executed. Then, we implement a strategy by the following specific steps.

*4.2.1 RMS*

Random Mutant Selection (RMS) selects a specified number or proportion of mutants from all mutants randomly. To implement RMS, given the number of selected mutants *n*, we keep randomly selecting one of the remaining mutants with equal probability until there are n mutants selected.

*4.2.2 Sentinel*

Sentinel is a state-of-the-art multi-objective evolutionary hyper-heuristic approach on mutant reduction [8]. For the "multi-objective evolutionary", it has two objective functions. In nature, the first objective is to maximize the average "absolute" strategy effectiveness of s over n runs (as there are randomness in s):

$$\uparrow \text{SCORE}(s) = \frac{\frac{1}{n}\sum_{i=1}^{n} MS(M,T_s(i))}{MS(M,T)} = \frac{1}{n}\sum_{i=1}^{n} \frac{|KM(M,T_s(i))|}{|KM(M,T)|}$$

As the |*KM*(*M, T*)| can be regard as a constant, this objective can be regard as maximizing the mutant killing ability of the test sets $T_s(i)$, $i = 1, \ldots, n$. The second objective is to minimize the execution time. Instead of acting over the search space of mutants, Sentinel searches the heuristic space for good mutant reduction strategies that can achieve the best score of objective functions on a training set. With several pre-defined basic mutant selection strategies (i.e. Random Mutant Selection, Random Operator Selection, and Certain Operator Selection), Sentinel searches a combination of basic strategies (e.g. randomly select two mutators to generate mutants and then select 10 mutants randomly among them) which can achieve the best values of the objective functions. After that, we can use the combination of basic strategies for selecting mutants on a target project. In [8], Guizzo et al. focused on cross-version mutant reduction. However, it is laborious to find a suitable training project for a target project. On the other hand, cross-version information for Sentinel is unfair to other strategies. As a result, we use the example project *Triangle* provided in the Sentinel's source code as the training data. At a high level, Sentinel provides a combined reduction strategy by searching the strategy space from the training project, which considers the execution cost and the "absolute" strategy effectiveness $E_s$ at the same time.

In our study, in order to implement Sentinel, given the number of selected mutants *n*, it can be used as a hyper parameter for training of Sentinel. That is to say, by setting this hyper parameter for the out-of-the-box Sentinel and after the training process, Sentinel can provide a combination of basic strategies that selects n mutants on the training program *Triangle*. Then, for a target program, we apply the combined strategy to the target program to select *n* mutants. It is worth noting that if the number of remaining mutants is going to be less than *n* after a basic strategy is executed, we will skip the basic strategy. For example, we set *n* = 4 and obtain the combined reduction strategy on Triangle: first, remove the mutator which will generate the most mutants; second, generate the mutants; third, select 4 mutants randomly. When we use this strategy on a target mutant set generated by only one mutator, none of the mutant will be generated after removing "the mutator which will generate the most mutants". As a result, we will skip this basic strategy and select 4 mutants randomly. We will discuss more details of Sentinel in Section 6.

*4.2.3 COS*

Certain Operator Selection (COS) selects the mutants generated by a specific mutator subset of all mutators. The 5-Operators Selection proposed by Offutt et al. [4] is the approach chosen in this paper. Table 3 summarizes the 5 mutation operators (shown in red color) selected from 22 mutation operators with the description. The 5 operators include the relational, logical, arithmetic, absolute, and unary insertion operators. For PIT, we remove *Return Values Mutator* and *Void Method Calls Mutator* among the 7 default mutators (i.e. *Conditionals Boundary Mutator,*

Table 4. The implement of COS for PIT

| Mutator In PIT | Type |
|---|---|
| ConditionalsBoundaryMutator | ROR |
| IncrementsMutator | AOR |
| InvertNegsMutator | UOI,ABS |
| MathMutator | AOR,LCR |
| NegateConditionalsMutator | ROR |
| ReturnValsMutator | |
| VoidMethodCallMutator | |

*Increments Mutator, Invert Negatives Mutator, Math Mutator, Negate Conditionals Mutator, Return Values Mutator, and Void Method Calls Mutator*) as the implementation of COS (as shown in Table 4). It is worth noting that these 7 default operators were called "defaults" mutators in the paper of Sentinel and we hence reuse this abbreviation. However, on the current PIT website [34], they are called "old-defaults" mutators. We report the name of each mutator with type, which is convenient for readers to reproduce and compare.

In our study, in order to implement COS, given the number of selected mutants $n$, we first delete the mutants generated by *Return Values Mutator* and *Void Method Calls Mutator*. Then, we keep randomly selecting one of the remaining mutants with equal probability until there are $n$ mutants selected. It is worth noting that COS can also make the remained number of mutants less than $n$. If such a situation happens, we will exclude the COS strategy from subsequent analysis.

*4.2.4 SMS*

Subsuming Mutant Selection (SMS) selects the subsuming mutants. By the definition in [13], "One mutant subsumes another if at least one test kills the first and every test that kills the first also kills the second". For all mutants, killing the subsuming mutants is to kill all the killable mutants. Therefore, the subsumed mutants should be regarded as redundancy. In the following, we first introduce the relevant definitions and then elaborate on the implementation of SMS in our study.

**Definition 1 (Subsuming).** Let $m_i$ and $m_j$ be two mutants in $M$. We say that $m_i$ subsumes $m_j$, if: (1) there exists some test cases kill $m_i$; and (2) for any test case in (1), it kills $m_j$.

**Definition 2 (Non-subsumed set).** The non-subsumed set is a subset of $M$. $m_i$ is an element of non-subsumed set if and only if none of the other mutants among $M$ subsumes $m_i$.

The goal of SMS is to find this non-subsumed set as the result of mutant selection. However, in order to determine the exact subsuming relationship, all mutants must be executed against all test cases, which is contrary to aim of mutant reduction. As a result, how to approximate the kill relationship matrix between test cases and mutants (the matrix in Fig. 1 is an example) is the main problem of SMS. In our study, we run the test suite on the original program SUT to obtain a coverage relationship matrix as the approximation of the kill relationship matrix. There are two reasons for us to take this as the approximation: on the one hand, covering is a necessary condition for killing, which reveals that the two matrixes are highly correlated; on the other hand, once the statement to be mutated is covered in SUT, the mutant will be covered (we say a test covers a mutant if and only if the test covers the corresponding statement to be mutated in the original program). As a result, the one time running for the test suite against SUT can obtain the coverage relationship matrix between test cases and mutants. This is a low-cost approach to approximate the kill relationship matrix. As such, we define the coverage subsuming relationship between two mutants as follows:

**Definition 3 (Coverage subsuming).** Let $m_i$ and $m_j$ be two mutants in $M$. We say that $m_i$ coverage-subsumes $m_j$, if: (1) there exists some test cases cover $m_i$; and (2) for any test case in (1), it covers $m_j$.

**Definition 4 (Non-coverage-subsumed set).** The non-coverage-subsumed set is a subset of $M$. $m_i$ is an element of non-coverage-subsumed set if and only if none of the other mutants among $M$ coverage-subsumes $m_i$.

Consequently, in our study, we define the Coverage-based Subsuming Mutant Selection to find this non-subsumed set as the result of mutant selection. In other words, we use the Coverage-based Subsuming Mutant Selection as the implementation of SMS. Specifically, Coverage-based Subsuming Mutant Selection consists of the following steps:

- Step 1: execute the test suite on SUT to collect the coverage information of the mutated statements corresponding to the mutants. By this step, a coverage relationship matrix is obtained;
- Step 2: find non-coverage-subsumed set as the selected mutants.

It is worth noting that SMS can only select a specific number of mutants, which cannot be given in advance.

*4.2.5 CMS*

Clustering Mutant Selection (CMS) [24] selects one mutant from each mutant cluster which is proposed by Hussain [24]. In [24], first, all mutants were executed against all test cases to obtain a kill relationship matrix. Each row of the kill relationship matrix was an instance corresponding to a mutation. For example, in Fig. 1, the instance of m1 is [1, 0, 0, 0]. Then, a clustering algorithm was applied on all mutant instances to classify the mutants into different clusters. Among each cluster, the mutants were guaranteed to be killed by similar test cases. Finally, the selection was obtained by randomly selecting one mutant from each cluster.

Since the kill relationship matrix is not available before executing all mutants, in our study, we use the coverage relationship matrix as the approximation. Specifically, in order to implement SMS, we use Coverage-based Clustering Mutant Selection consists with the following steps in this paper:

- Step 1: execute the test suite on SUT to collect a coverage information of the mutated statements corresponding to the mutants. By this step, the coverage relationship matrix is obtained;
- Step 2: given the number of selected mutants n, a k-means algorithm is used to obtain n clusters by mutant instances from the coverage relationship matrix;
- Step 3: randomly select one mutant from each cluster to obtain the selected mutants.

**4.3 Research questions**

The RQs mainly focus on two aims: one is to investigate the distinguishing ability of mutation reduction evaluation indicators, while the other is to find the effective mutation reduction strategies. However, the two aims are not independent. First, we believe that there are differences between these reduction strategies. If an indicator makes all strategies look the same numerically on most programs, it has a weak discrimination ability. If an indicator makes some strategies ahead of the others numerically on most programs, it has a strong discrimination ability. On the whole, the two aims are achieved at the same time: by comparing the five strategies with different indicators, we will find the indicators with a strong discrimination ability as well as the effective strategies. It is worth noting that we only compare OP with variation of mutation score (*VMS*), as the irrationality of the *strategy effectiveness* $E_s$ has been fully clarified (i.e. in Section 3.2.2, a useless reduction strategy can achieve the best strategy effectiveness). For EROP, on the one hand, we must examine the rationality of OP before verifying EROP. On the other hand, EROP is the first indicator which considers the number of reductions and the effectiveness of mutation reduction at the same time. As a result, we put EROP in the last RQ to investigate the effectiveness of reduction strategies.

*RQ1 (Discrimination ability under a sufficient test suite): Which is more reasonable as the indicator to evaluate reduction strategy, the variation of Mutation Score (VMS) or Order Preservation (OP), under a sufficient test suite?*

The purpose of RQ1 is to investigate which evaluation indicator can find the best reduction strategy under a sufficient test suite. In our study, each of the subject programs has a relatively sufficient test suite. In order to study RQ1, we use 10 programs with their test suites (see Section 4.1 for detail) directly for the experiment without any change. In order to investigate RQ1, we apply the 5 strategies to 10 programs for 100 times and calculate *VMS* and OP against the sufficient test suites. The basic idea is that these strategies should not be difficult to distinguish by a sensitive indicator, and effective strategies will stand out on most projects. To see the difference, on the one hand, we list the average results for 10 programs. On the other hand, we draw the boxplots for each strategy with 100 *VMS* and 100 OPs against each program.

Note that SMS can only provide an unchangeable number of selected mutants. As a result, under RQ1, for each of the other strategies, we set the same number of selected mutants as the number of mutants selected by SMS. In particular, for CMS, we first implement SMS and then use the selection result as the initial clustering center of CMS. This is to make the initial cluster centers as different as possible.

*RQ2 (Discrimination ability under an insufficient test suite): Which is more reasonable as the indicator to evaluate reduction strategy, the variation of Mutation Score (VMS) or Order Preservation (OP), under an insufficient test suite?*

The purpose of RQ2 is to investigate whether the result of RQ1 will change or not under an insufficient test suite. In the literature, few studies use insufficient test suites to investigate the effectiveness of mutation reduction strategies. Combining the results from sufficient and insufficient test suites, we will get an in-depth understanding the usefulness of our proposed mutation reduction indicators. In order to simulate an insufficient test suite, we first delete test cases used in RQ1 with even index for each program. After that, for RQ2, we apply the 5 strategies to 10 programs for 100 times and calculate *VMS* and OP against the insufficient test suites. To see the difference, on the one hand, we list the average results for 10 programs. On the other hand, we draw the boxplots for each strategy with 100 *VMS* and 100 OPs against each program. Note that, we take the same method as used in RQ1 to set the number of selected mutants and to implement CMS.

*RQ3 (Influence of reduction ratio on reduction effectiveness): How do the effectiveness of reduction strategies in terms of OP change under different mutation reduction ratios?*

The purpose of RQ3 is to investigate the influence of mutation reduction ratio on reduction effectiveness. In the process of exploring this question, on the one hand, we can have a more comprehensive understanding of mutation reduction strategies, on the other hand, we can know how much mutation reduction is the best. To our knowledge, quantitative analysis of reduction ratio is scarce. To answer this question, we set the reduction ratio from 90% to 10% with a step length of 10% for J1 while setting the reduction ratio from 95% to 0% with a step length of 5% for other programs. For each ratio, we first apply the feasible strategies to 10 programs for 100 times and calculate an average OP and an average EROP against the sufficient test suites. Then, with 19 OP values and EROP values under different ratios, the OP-Reduction Ratio scatter diagram and EROP-Reduction Ratio scatter diagram of each strategy can be drawn. It is worth noting that SMS can only be reduced to a fixed number. As a result, we also add a group of 5 points under the specific reduction ratio (as shown in the last column in Table 5) for showing SMS. Besides,

Table 5. The discrimination ability comparison between OP and *VMS* under a sufficient test suite

(the strategy rank by the Scott-Knott ESD test is shown in brackets and a lower rank is better)

| Project | OP | | | | | Variation of MS | | | | | Reduction Ratio |
|---|---|---|---|---|---|---|---|---|---|---|---|
| | RMS | Sentinel | COS | SMS | CMS | RMS | Sentinel | COS | SMS | CMS | |
| J1 | 0.615 (3) | 0.270 (4) | 0.270 (4) | 0.872 (1) | 0.777 (2) | 0 (1) | 0 (1) | 0 (1) | 0 (1) | 0 (1) | 7/11 |
| J2 | 0.913 (3) | 0.920 (3) | - | 0.956 (2) | 0.970 (1) | 0.023 (1) | 0.020 (1) | - | 0.026 (2) | 0.026 (2) | 18/39 |
| J3 | 0.403 (4) | 0.475 (3) | 0.313 (5) | 0.754 (1) | 0.697 (2) | 0.170 (2) | 0.158 (2) | 0.163 (2) | 0.161 (2) | 0.148 (1) | 58/62 |
| J4 | 0.893 (2) | 0.851 (3) | - | 0.891 (2) | 0.936 (1) | 0.045 (1) | 0.092 (2) | - | 0.119 (3) | 0.046 (1) | 88/109 |
| J5 | 0.698 (3) | 0.596 (4) | 0.586 (5) | 0.708 (2) | 0.724 (1) | 0.071 (2) | 0.296 (5) | 0.284 (4) | 0.075 (2) | 0.058 (1) | 96/115 |
| J6 | 0.616 (3) | 0.513 (5) | 0.574 (4) | 0.700 (2) | 0.754 (1) | 0.104 (1) | 0.173 (3) | 0.110 (1) | 0.200 (4) | 0.144 (2) | 200/213 |
| J7 | 0.620 (3) | 0.623 (3) | 0.604 (4) | 0.658 (2) | 0.699 (1) | 0.123 (3) | 0.130 (3) | 0.099 (2) | 0.013 (1) | 0.112 (2) | 141/149 |
| J8 | 0.525 (1) | 0.528 (1) | 0.516 (2) | 0.531 (1) | 0.496 (3) | 0.063 (1) | 0.115 (2) | 0.081 (1) | 0.463 (4) | 0.161 (3) | 213/215 |
| J9 | 0.649 (3) | 0.594 (4) | 0.654 (3) | 0.839 (1) | 0.749 (2) | 0.111 (2) | 0.175 (3) | 0.116 (2) | 0.093 (1) | 0.106 (2) | 126/134 |
| J10 | 0.879 (4) | 0.892 (3) | 0.865 (5) | 0.947 (2) | 0.972 (1) | 0.023 (1) | 0.077 (4) | 0.026 (1) | 0.058 (3) | 0.036 (2) | 1103/1161 |
| avg. | 0.681 | 0.626 | 0.547 | 0.787 | 0.778 | 0.073 | 0.126 | 0.110 | 0.121 | 0.084 | - |

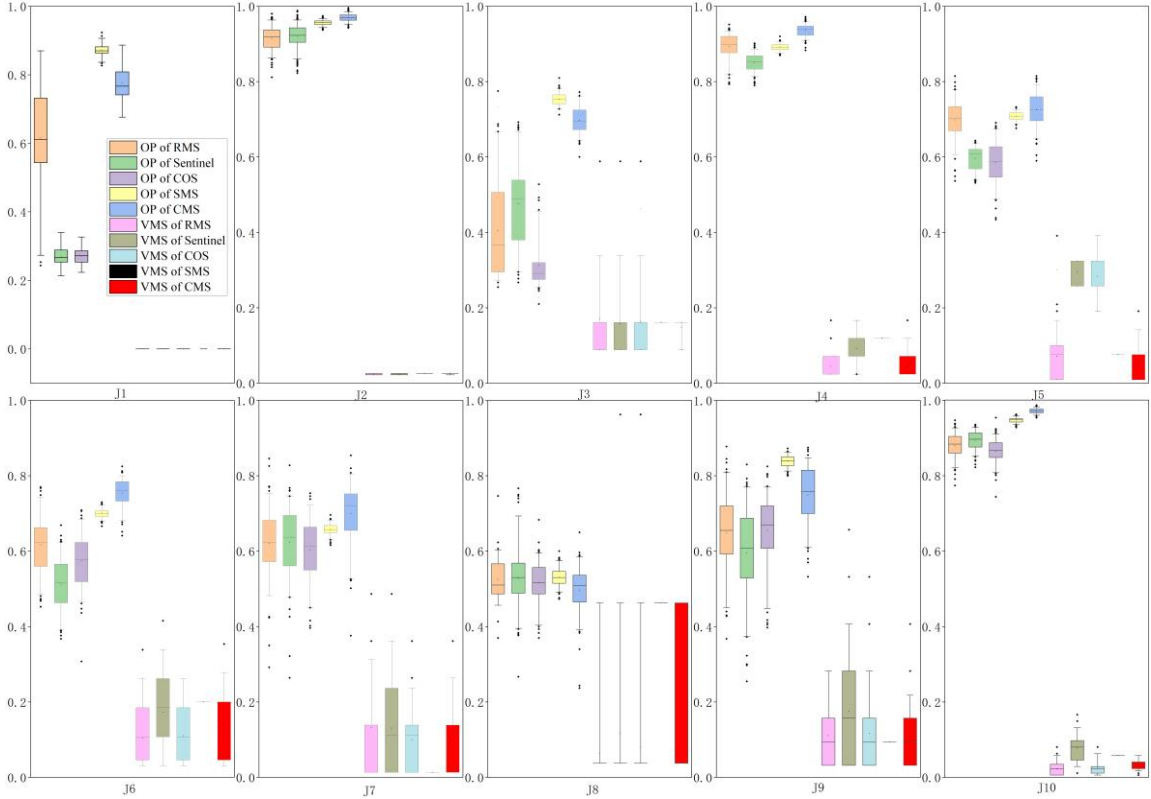

Figure 3. The discrimination ability comparison between OP and *VMS* under a sufficient test suite. The legends from top to down are OP of RMS, Sentinel, COS, SMS, and CMS, *VMS* of RMS, Sentinel, COS, SMS, and CMS, respectively. Note that COS is skipped in J2 and J4 since there are less than 21 (39-18 and 109-88) mutants after filtering two mutators which means COS cannot be set to the same reduction ratio as SMS.

COS has only a few data points due to the pre-deletion of some mutants. For CMS under RQ3, we use the default settings of *sklearn*[3] to generate the initial cluster centers. The k of k-means is set to the number of selected mutants in RQ1. For example, if there are 4 mutants selected in J1 under RQ1, CMS will first cluster mutants into 4 clusters for J1. Then, CMS randomly selects one mutant from each cluster in turn until x% mutants are selected.

## 5 EXPERIMENTAL RESULTS

In this section, we report in detail the experimental results on the ability of OP and EROP to distinguish mutation reduction effectiveness under sufficient and insufficient test suites, as well as the influence of mutation reduction ratio on their discrimination ability.

---

[3] https://scikit-learn.org/stable/modules/generated/sklearn.cluster.KMeans.html.

## 5.1 RQ1: Discrimination ability under a sufficient test suite

Table 5 reports for 5 strategies the average OP and *VMS* values over 100 times reduction results on each program. The EROP is not shown in RQ1 and RQ2, which can be computed from the data in the table. In particular, for each of OP and *VMS*, we applied the improved Scott-Knott Effect Size Difference (ESD) test (v 2.0) [39] to compute for each strategy a rank (as shown in brackets, a lower rank is better). The Scott-Knott ESD test clustered variables (the OPs or *VMS*s of 5 strategies in our context) into different groups to obtain their ranks according to statistically significant differences in their mean variable importance scores ($\alpha = 0.05$). During this process, two groups were merged when a pair of statistically distinct groups had a negligible Cohen's d effect size [40]. By convention, $|d| < 0.2$ means "negligible", $0.2 \leq |d| < 0.5$ means "small", $0.5 \leq |d| < 0.8$ means "medium", and $|d| \geq 0.8$ means "large". Fig. 3 shows for each of 5 strategies the boxplot over 100 times mutation reductions. From Table 5 and Fig. 3, we have the following observations:

- From the result on J1, we can find that, if the mutation score of a program is 100%, any selected mutant set will achieve a 100% mutation score. In other words, using *VMS* as a reduction effectiveness evaluation indicator is useless and misleading which makes all the evaluations the same. However, OP can clearly distinguish several strategies.

- By OP, SMS and CMS are more outstanding than RMS. In Table 5, on average, SMS and CMS have a 15% higher OP than RMS while Sentinel and COS have lower OP than RMS. In Fig. 3, the boxes of SMS and CMS are obviously higher than the other mutation reduction strategies on most programs. On the whole, it is easy to distinguish several strategies. We speculate that the reasons for the poor performance of Sentinel are as two-fold: (1) it is originally designed for cross-version prediction scenario and is limited in cross-project prediction scenario; and (2) since its objective function is not on OP, it may be difficult to achieve a good performance under OP.

- By *VMS*, Sentinel, COS, and SMS are the "bad" strategies, while RMS is the "best" strategies. In Table 5, most of the average variations fluctuate around 0.1. However, the gap between projects is large. For J5, Sentinel and COS achieve two times worse *VMS* than their average *VMS* while SMS and CMS achieve two times worse *VMS* for J8. In Fig. 3, since SMS always selects the same mutants without randomness, the box for SMS is a line. For the other strategies, there is little difference on J2, J3 J4, J9, and J10. On the whole, it is not easy to distinguish several strategies since most are similar except Sentinel looks a little worse. For RMS, as analyzed in 3.2.1, RMS can maintain the mutation score in mathematical expectation, which can explain the best performance under *VMS*. This observation is consistent with many prior conclusions: many strategies performed close to RMS [20] [27]. However, we use this observation stand for the poor discrimination ability of *VMS* instead of standing for these conclusions.

- By the rank, on the one hand, the rank of OP can cluster strategies into more groups than *VMS* on most programs. This observation reveals the fact that OP has a stronger discrimination ability than *VMS*. On the other hand, the ranking result of OP is more consistent than *VMS* across projects: for OP, SMS and CMS can achieve the lower rank on most programs than the other strategies; for *VMS*, Sentinel, COS, SMS and CMS can achieve the highest (worst) rank on some programs while achieving the lowest (best) rank on some other programs. To sum the two observations, not only the discrimination ability of OP is stronger than *VMS*, but also the confidence of OP's conclusions is higher than *VMS*'s.

Table 6. The discrimination ability comparison between OP and Variation of *MS* under an insufficient test suite
(the strategy rank by the Scott-Knott ESD test is shown in brackets and a lower rank is better)

| Project | OP | | | | | Variation of MS | | | | | Reduction Ratio |
|---|---|---|---|---|---|---|---|---|---|---|---|
| | RMS | Sentinel | COS | SMS | CMS | RMS | Sentinel | COS | SMS | CMS | |
| J1 | 0.725 (3) | 0.720 (3) | - | 1.000 (1) | 0.875 (2) | 0.097 (2) | 0.107 (2) | - | 0.018 (1) | 0.120 (3) | 6/11 |
| J2 | 0.764 (2) | 0.753 (2) | - | 0.881 (1) | 0.882 (1) | 0.084 (2) | 0.077 (1) | - | 0.085 (2) | 0.102 (3) | 27/39 |
| J3 | 0.433 (3) | 0.453 (3) | 0.393 (4) | 0.820 (1) | 0.599 (2) | 0.200 (2) | 0.245 (3) | 0.180 (1) | 0.210 (2) | 0.216 (2) | 59/62 |
| J4 | 0.813 (2) | 0.791 (3) | - | 0.814 (2) | 0.895 (1) | 0.075 (2) | 0.098 (3) | - | 0.058 (1) | 0.079 (2) | 95/109 |
| J5 | 0.688 (3) | 0.534 (4) | 0.536 (4) | 0.724 (1) | 0.780 (1) | 0.091 (1) | 0.302 (3) | 0.306 (3) | 0.118 (2) | 0.099 (1) | 100/115 |
| J6 | 0.499 (3) | 0.410 (5) | 0.479 (4) | 0.737 (1) | 0.651 (2) | 0.126 (1) | 0.170 (2) | 0.127 (1) | 0.409 (3) | 0.141 (1) | 203/213 |
| J7 | 0.535 (3) | 0.547 (3) | 0.539 (3) | 0.756 (1) | 0.698 (2) | 0.152 (2) | 0.129 (1) | 0.133 (1) | 0.181 (3) | 0.181 (3) | 142/149 |
| J8 | 0.646 (2) | 0.641 (2) | 0.666 (1) | 0.631 (3) | 0.608 (4) | 0.087 (1) | 0.162 (3) | 0.116 (2) | 0.453 (5) | 0.214 (4) | 213/215 |
| J9 | 0.643 (3) | 0.595 (4) | 0.636 (3) | 0.862 (1) | 0.760 (2) | 0.119 (1) | 0.130 (2) | 0.114 (1) | 0.239 (3) | 0.107 (1) | 126/134 |
| J10 | 0.779 (4) | 0.793 (3) | 0.754 (5) | 0.931 (2) | 0.962 (1) | 0.042 (1) | 0.095 (3) | 0.047 (1) | 0.125 (4) | 0.074 (2) | 1128/1161 |
| avg. | 0.653 | 0.624 | 0.572 | 0.816 | 0.771 | 0.107 | 0.145 | 0.146 | 0.182 | 0.129 | - |

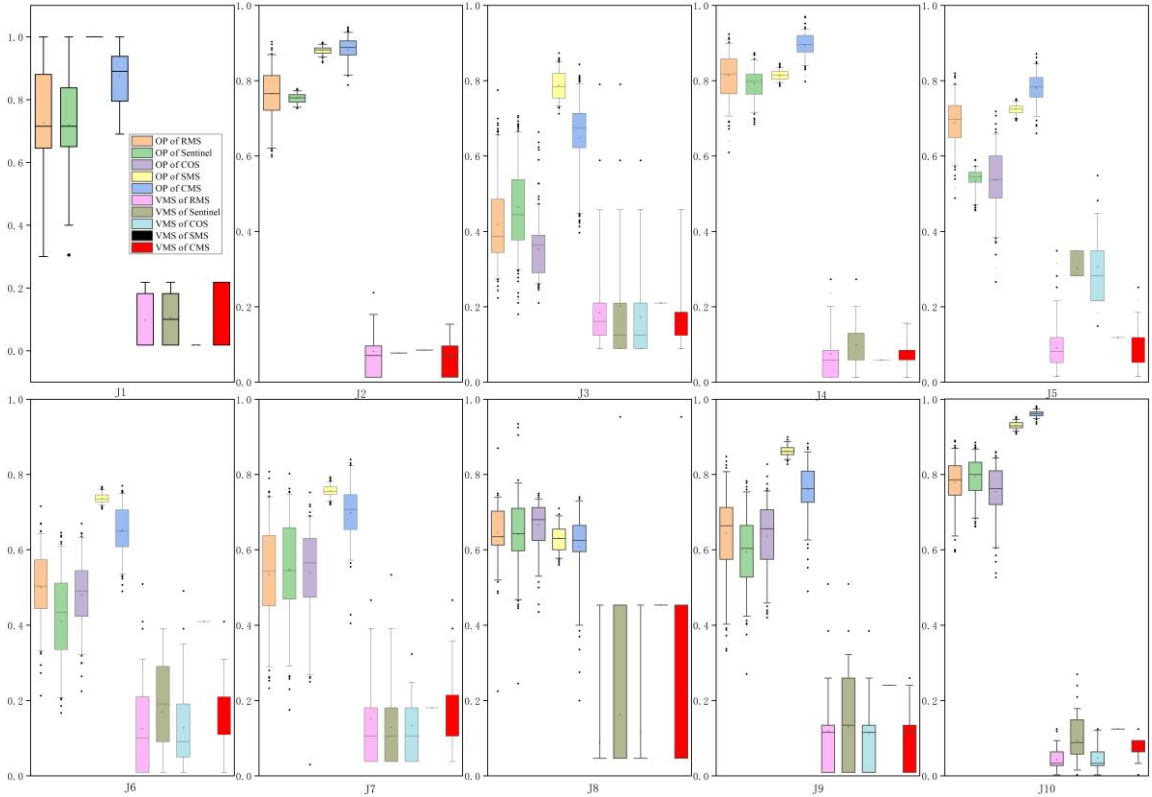

Figure 4. The discrimination ability comparison between OP and *VMS* under an insufficient test suite. The legends from top to down are OP of RMS, Sentinel, COS, SMS, and CMS, *VMS* of RMS, Sentinel, COS, SMS, and CMS, respectively. It is worth noting that COS is skipped in J1, J2, and J4 since there are respectively less than 5,12, and 14 mutants after filtering two mutators, which means COS cannot be set to the same reduction ratio as SMS.

The conclusion obtained by OP is inconsistent to the conclusion obtained by *VMS*. Combining the above observations with the previous theoretical analysis, we can conclude that the use of OP is more objective and accurate, and the use of *VMS* may lead to misleading conclusions.

> To conclude, under sufficient test suites, OP is a better indicator than *VMS* in mutation reduction evaluation, while SMS and CMS are more effective than RMS, Sentinel, and COS under OP.

**5.2 RQ2: Discrimination ability under an insufficient test suite**

Table 6 reports for 5 strategies the average OP and *VMS* values over 100 times reduction results on each program. Fig. 4 shows for each of the 5 strategies the boxplot over 100 times mutation selections. From Table 6 and Fig. 4, we have the following observations:

- By OP, SMS and CMS are more outstanding than the other strategies. As shown in Table 6, on average, SMS has a 25% higher OP than RMS, CMS has a 20% higher OP than RMS, and Sentinel and COS have a lower

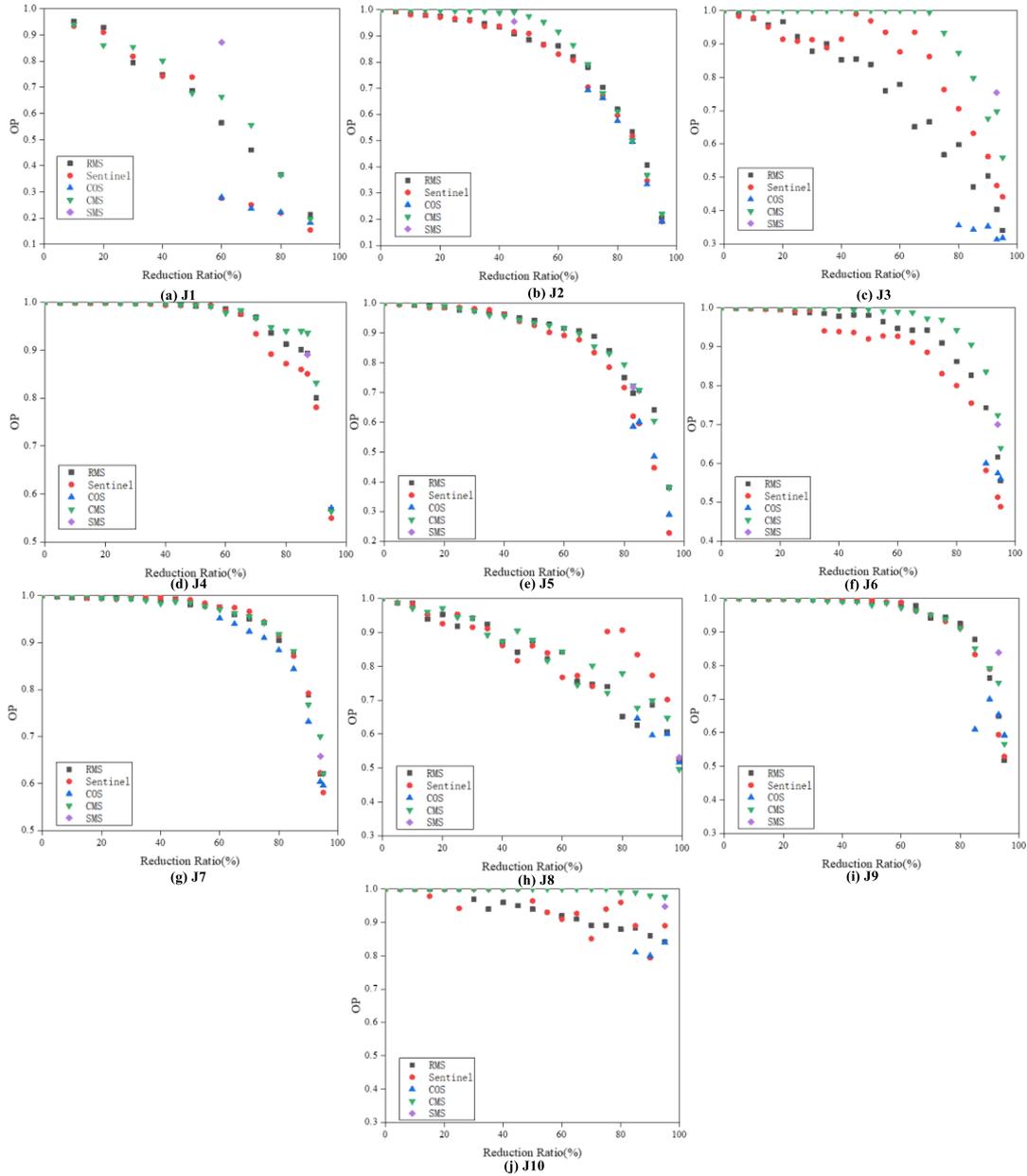

Figure 5. The OP-reduction ratio scatter diagram.

OP than RMS. As shown in Fig. 4, the boxes of SMS and CMS are obviously higher than the others on most programs. On the whole, it is easy to distinguish several strategies under OP. Overall, this conclusion is consistent with RQ1.

- By *VMS*, SMS is the "worst" strategy while RMS is still the "best" strategy. SMS achieves the worse *VMS* because of the highest *VMS* on J6 and J8. As shown in Table 6 and Fig. 4, there is little difference for most of the 5 strategies on J1, J2, J3, J7, J8, and J9. Moreover, for some programs, different strategies have a similar *VMS* distribution (e.g., RMS and Sentinel on J1). On the whole, it is not easy to distinguish several strategies since except SMS looks a little worse on average while the other methods are similar. The difference to conclusion on *VMS* in RQ1 is that the "worst" strategy changes from Sentinel to SMS, and the other strategies still seem difficult to distinguish.

- By the rank, on the one hand, the rank of OP can cluster strategies into more groups than *VMS* on 5 programs while clustering strategies into same number of groups to *VMS* on 4 programs. On the other hand, the ranking result of OP is more consistent than *VMS* across projects: for OP, SMS and CMS can achieve a lower rank on most programs than the other strategies; for *VMS*, Sentinel, COS, SMS, and CMS can achieve the highest (worst) rank on some programs while achieving the lowest (best) rank on some other programs. Compared with RQ1, although the groups by OP is less than that in RQ1, the variance of OP's rank is not larger than 1 (e.g., on J1, the variance of OP's rank for Sentinel is $|4-3|=1$) which means the ranks of OP are stable on sufficient and insufficient test suites. However, the variance of *VMS*'s rank maybe 2 (e.g., on J7, the variance of OP's rank for Sentinel is $|3-1|=2$) on some programs. Considering that there are only 3 groups under *VMS* on 8 out of 10 programs, these variance is relatively large. As a result, the ranks of *VMS* are erratic on different test suites. To sum these observations, OP not only has a stronger discrimination ability but also can lead to conclusions with a higher confidence compared with *VMS*.

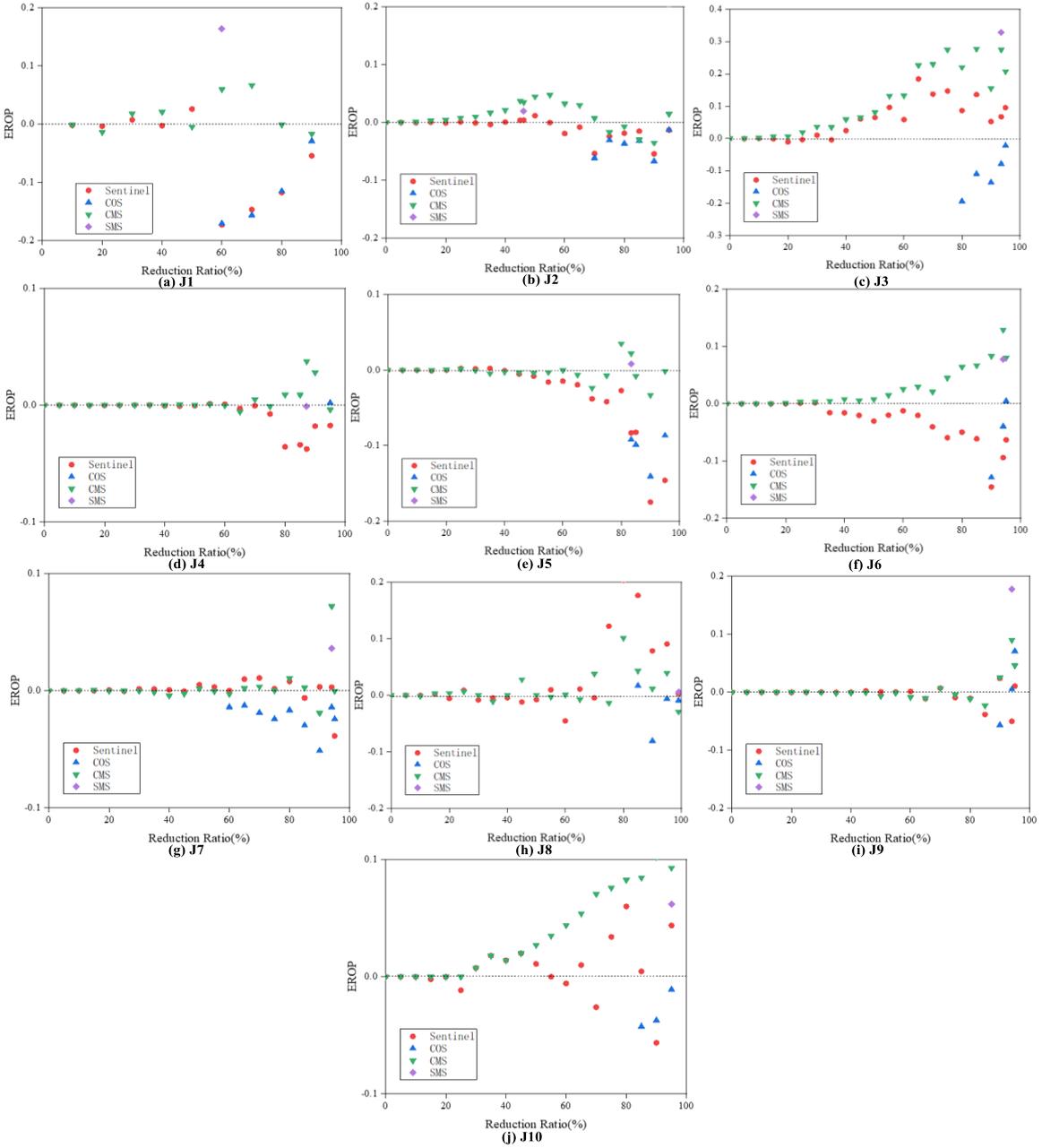

Figure 6. The EROP-reduction ratio scatter diagram.

- Another interesting finding is that *VMS* has increased compared with RQ1. In the literature [4,6-11], many strategies were reported have a small *VMS* under sufficient test suites and hence were regarded as effective in mutation reduction. However, our finding reveals that the conclusions drawn from sufficient test suites may not be generalized to insufficient test suites.

The main conclusions in RQ2 are consistent with that in RQ1: (1) OP has a stronger discrimination ability compared with *VMS*; (2) SMS and CMS are more effective than RMS, COS, and Sentinel under OP; and (3) it is still hard to distinguish 5 strategies by *VMS*.

> To conclude, under insufficient test suites, OP is still a better indicator than *VMS* in mutation reduction evaluation. In particular, the conclusions drawn from OP are stable when test suites change from sufficient to insufficient. However, whether test suites are sufficient or not may have a large influence on the conclusions drawn from *VMS*.

**5.3 RQ3: Influence of reduction ratio on reduction effectiveness**

Fig. 5 shows the OP-reduction ratio scatter diagrams on 10 programs. In this subsection, we focus on 4 feasible strategies since SMS has been completely compared in RQ1 and RQ2. From Fig. 5, we have the following observations:

- For all the mutation reduction strategies, on the whole, a higher reduction ratio leads to a greater loss in OP (i.e., the order preservation ability). Furthermore, for the 4 feasible strategies, setting the reduction ratio to 0.4 (i.e., select 60% mutants among all mutants) can achieve an OP value larger than 0.9 on most programs. Considering that CMS need coverage relationship matrix for implementation, when practitioners focus on maintaining the relative evaluation of test suites, selecting 60% mutants randomly is an easy choice than using CMS. Furthermore, we can see that existing reduction strategies may have an unsatisfactory OP (i.e. $< 0.9$) under a high reduction ratio (i.e. $> 0.8$). Therefore, in the future, a natural research problem is how to keep a high reduction ratio and a high OP at the same time?

- For the 4 feasible strategies, CMS shows the best reduction effectiveness on J1, J2, J3, J4, J6, and J10, which can maintain a relatively high OP and a relatively high reduction ratio at the same time. COS looks worse than RMS while Sentinel looks close to RMS on most programs. We speculate that PIT has retained most of the valuable mutation operators in the process of screening mutation operators, and hence it is of little meaning to select operators on this basis. For Sentinel, the essence of this approach is to bundle a lot of random strategies. As a result, Sentinel has a similar effectiveness compared with RMS.

Fig. 6 shows the EROP-reduction ratio scatter diagrams on 10 programs. From Fig. 6, we have the following observations:

- On the whole, setting a reduction ratio between 60% and 80% is both effective and efficient for CMS. A reduction ratio less than 60% has little effect for all strategies, while a reduction ratio more than 80% has unstable effects among different programs. For COS and Sentinel, it is hard to recommend a reduction ratio for all programs.

- CMS shows the best performance, which is above other strategies on the whole. Sentinel has a performance close to RMS while COS is the most ineffective strategy. Considering the unstable performance of COS and Sentinel on different programs, we suggest using RMS if it is hard to collect the coverage information for CMS.

Combining the above results, we can say that CMS is a more effective strategy.

> To conclude, for each reduction strategy, the reduction effectiveness as measured by OP decreases with the reduction ratio. For the best performing reduction strategy CMS, setting a reduction ratio between 60% and 80% is both effective and efficient. By EROP, it is difficult for existing strategies to significantly exceed RMS when the reduction ratio is less than 50%. When the reduction ratio is larger than 50%, only SMS and CMS can achieve a better performance than RMS on most programs.

# 6 DISCUSSION

In this section, we first discuss why not use the average variation of mutation scores on multiple test suites as the indicator to evaluate mutation reduction effectiveness. Then, we examine whether the subset constraint can be removed when generating a number of test suits to compute OP. After that, we diagnose whether the way to generate insufficient test suites affects the conclusions of RQ2. Finally, we analyze the reason why Sentinel has a low OP value.

## 6.1 Why not use the average variation of mutation scores on multiple test suites as the indicator?

In the current literature, it is a standard practice to use *VMS* from a single test suite to evaluate the effectiveness of a reduction strategy, i.e., the ability to maintain mutation score. However, as analyzed in Section 3, in order to obtain a comprehensive evaluation, we should assess the ability of a mutation reduction strategy to maintain mutation score on ALL test suites. In practice, it is not feasible to generate all possible test suites for mutation reduction evaluation. As a compromise, we can use multiple test suites to conduction the evaluation. In this context, a natural problem is: what if the average *VMS* on multiple test suites is used as the evaluation indicator? To figure out this question, we compute for each program the average *VMS* (as shown in Table 7) on the same test suites that are used to compute OP. Note that the settings here are consistent with RQ1. From Table 7, we have the following observations:

- Compared with Table 5, the average *VMS* on multiple test suites is larger than *VMS* on the original test suite for most strategies on most projects. Besides, the rank groups are more than that in Table 5, which means the discrimination ability of average *VMS* is stronger than *VMS*.
- RMS achieves the "best" performance according to average *VMS*. There is little difference among COS, Sentinel, and CMS on average. The large variations of SMS on J3, J4, J8, and J10 lead to the poor 'avg' result of 10 programs. The other cells for the five strategies are around 0.1~0.2.
- Compared with OP, it is not easy to distinguish these strategies by average *VMS*. Although average *VMS* produces more ranks than *VMS*, Sentinel, COS, SMS, and CMS can achieve the "1" or "2" rank on some programs while achieving the "4" or "5" rank on the other programs. As a result, it is hard to use average *VMS* to distinguish these reduction strategies based on the different ranks among 10 programs.

Based on the above observations, we can see that the average *VMS* still has an unsatisfactory ability to distinguish mutation reduction strategies. Furthermore, in nature, the average *VMS* still measures the ability to maintain mutation score rather than the "order-preserving ability". As a result, we do not recommend using the average *VMS* as the indicator for evaluation of the effectiveness of a reduction strategy.

Table 7. The discrimination ability comparison between OP and average *VMS* under a sufficient test suite (the strategy rank by the Scott-Knott ESD test in brackets and a lower rank is better)

| Project | OP | | | | | Average Variation of MS | | | | | Reduction Ratio |
|---|---|---|---|---|---|---|---|---|---|---|---|
| | RMS | Sentinel | COS | SMS | CMS | RMS | Sentinel | COS | SMS | CMS | |
| J1 | 0.615 (3) | 0.270 (4) | 0.270 (4) | 0.872 (1) | 0.777 (2) | 0.093 (1) | 0.140 (3) | 0.140 (3) | 0.125 (2) | 0.099 (1) | 7/11 |
| J2 | 0.913 (3) | 0.920 (3) | - | 0.956 (2) | 0.970 (1) | 0.050 (2) | 0.048 (2) | - | 0.066 (3) | 0.040 (1) | 18/39 |
| J3 | 0.403 (4) | 0.475 (3) | 0.313 (5) | 0.754 (1) | 0.697 (2) | 0.154 (1) | 0.197 (3) | 0.143 (1) | 0.374 (4) | 0.169 (2) | 58/62 |
| J4 | 0.893 (3) | 0.851 (3) | - | 0.891 (2) | 0.936 (1) | 0.069 (1) | 0.088 (2) | - | 0.216 (4) | 0.097 (3) | 88/109 |
| J5 | 0.698 (3) | 0.596 (4) | 0.586 (5) | 0.708 (2) | 0.724 (1) | 0.087 (1) | 0.185 (4) | 0.211 (5) | 0.160 (3) | 0.101 (2) | 96/115 |
| J6 | 0.616 (3) | 0.513 (5) | 0.574 (4) | 0.700 (2) | 0.754 (1) | 0.082 (2) | 0.127 (3) | 0.087 (2) | 0.068 (1) | 0.119 (3) | 200/213 |
| J7 | 0.620 (3) | 0.623 (3) | 0.604 (4) | 0.658 (2) | 0.699 (1) | 0.128 (3) | 0.115 (2) | 0.123 (3) | 0.104 (1) | 0.106 (1) | 141/149 |
| J8 | 0.525 (1) | 0.528 (1) | 0.516 (2) | 0.531 (1) | 0.496 (2) | 0.107 (1) | 0.172 (2) | 0.146 (3) | 0.527 (5) | 0.247 (4) | 213/215 |
| J9 | 0.649 (2) | 0.594 (4) | 0.654 (3) | 0.839 (1) | 0.749 (2) | 0.123 (2) | 0.176 (3) | 0.123 (2) | 0.168 (3) | 0.106 (1) | 126/134 |
| J10 | 0.879 (4) | 0.892 (3) | 0.865 (5) | 0.947 (2) | 0.972 (1) | 0.045 (1) | 0.090 (2) | 0.050 (1) | 0.348 (4) | 0.283 (3) | 1103/1161 |
| avg. | 0.681 | 0.626 | 0.547 | 0.787 | 0.778 | 0.094 | 0.134 | 0.128 | 0.216 | 0.137 | - |

## 6.2 Can the subset constraint be removed when generating test suits to compute OP?

In Section 3.3, given a SUT with the original test suite $T$, in order to compute OP, we employ a "continuous half-sample" approach to generate a sequence of test suites from $T$ such that they have a subset constraint. Specifically, the subset constraint enables to obtain a sequence of test suites by generating "$T_{i+1}$ by randomly selecting half of the test cases (round down) in $T_i$". In nature, this subset constraint places a restriction on which pairs of test suites the mutation score order can be used to compute OP. In other words, since $T_{i+1}$ is a subset of $T_i$, we know that $T_{i+1}$ must have a weaker or at most the same ability in detecting faults compared with $T_i$. In this context, the subset constraint ensures that the mutation score order on $(T_i, T_{i+1})$ before mutation reduction can be regarded as the ground truth in fault detecting potential. As a result, OP can reflect to what extent a reduction strategy maintains the ability to measure the effectiveness of test suites in detecting faults.

However, for most of the existing studies, the mutation score is regarded as the ground truth among all the test suites. In other words, given two test suites, regardless of whether two suites have a subset relation, the test suite with a higher mutation score is considered more effective in detecting faults than the other test suite. Under this assumption, a natural problem is immediately raised: will the conclusions of OP be changed? To investigate this problem, in this subsection, we remove the subset constraint to compare two suites based on such an assumption: the mutation score order here is indeed a set of ($T1$, $T2$, operator) satisfying operator is ">" if $T1$ has a higher mutation score and is "=" if $T1$ has the same mutation score compared with $T2$. By this definition, we can judge the effectiveness relationship between any two test suites. To compare with the original OP, we first randomly select 100*k (k = int($\log_2|T_0|$), see Section 3.3 to find the detail of k) pairs of test suites among the subset space of the original test suite $T_0$ to compute a New OP (NOP). If the operator on ($T1$, $T2$, operator) is not changed before and after reduction., we refer to ($T1$, $T2$) to an order-preserving pair. Then, we divide the total number of order-preserving pairs by 100*k to obtain NOP.

Table 8 reports for each program the OP and NOP values under five mutation reduction strategies. Note that the settings here are consistent with RQ1. From Table 8, we have the following observations:

- By NOP, the conclusion on five reduction strategies is consistent with RQ1: SMS and CMS are better than RMS, Sentinel, and COS. The difference between SMS and CMS is trivial.

- Overall, NOP is larger than OP. We do not have enough evidence to prove whether NOP overestimates the "order-preserving ability" or OP underestimates it. The cause for this phenomenon is not clear, which is a direction of future work.

However, we do not recommend using this approach to calculate of the "order-preserving ability" of a reduction strategy, since the assumption may be questionable. Papadakis et al. concluded that "MS might not be a reliable indicator of the test effectiveness" [38]. Compared with NOP, the effectiveness relationship of the test suite sequence used by OP is a ground truth based on *score monotonicity property* (see footnote in 3.3 for detail). As a result, we believe that the confidence of OP is stronger than NOP. To conclude, the conclusion of NOP is consistent with the previous conclusion. It is possible to use NOP in future studies, but we do not recommend it since NOP requires stronger assumptions than OP.

Table 8. The discrimination ability comparison between OP and NOP under a sufficient test suite
(the strategy rank by the Scott-Knott ESD test is shown in brackets and a lower rank is better)

| Project | OP | | | | | NOP | | | | | Reduction Ratio |
|---|---|---|---|---|---|---|---|---|---|---|---|
| | RMS | Sentinel | COS | SMS | CMS | RMS | Sentinel | COS | SMS | CMS | |
| J1 | 0.615 (3) | 0.270 (4) | 0.270 (4) | 0.872 (1) | 0.777 (2) | 0.676 (3) | 0.435 (4) | 0.363 (5) | 0.860 (1) | 0.790 (2) | 7/11 |
| J2 | 0.913 (3) | 0.920 (3) | - | 0.956 (2) | 0.970 (1) | 0.891 (4) | 0.905 (3) | - | 0.917 (2) | 0.941 (1) | 18/39 |
| J3 | 0.403 (4) | 0.475 (3) | 0.313 (5) | 0.754 (1) | 0.697 (2) | 0.368 (3) | 0.286 (4) | 0.235 (5) | 0.880 (1) | 0.805 (2) | 58/62 |
| J4 | 0.893 (2) | 0.851 (3) | - | 0.891 (2) | 0.936 (1) | 0.801 (4) | 0.843 (3) | - | 0.878 (2) | 0.914 (1) | 88/109 |
| J5 | 0.698 (3) | 0.596 (4) | 0.586 (5) | 0.708 (2) | 0.724 (1) | 0.775 (3) | 0.755 (4) | 0.716 (5) | 0.880 (2) | 0.904 (1) | 96/115 |
| J6 | 0.616 (3) | 0.513 (5) | 0.574 (4) | 0.700 (2) | 0.754 (1) | 0.778 (3) | 0.737 (5) | 0.759 (4) | 0.888 (2) | 0.899 (1) | 200/213 |
| J7 | 0.620 (3) | 0.623 (3) | 0.604 (4) | 0.658 (2) | 0.699 (1) | 0.706 (4) | 0.731 (3) | 0.654 (5) | 0.820 (2) | 0.845 (1) | 141/149 |
| J8 | 0.525 (1) | 0.528 (1) | 0.516 (2) | 0.531 (1) | 0.496 (3) | 0.583 (3) | 0.595 (2) | 0.560 (4) | 0.620 (1) | 0.593 (2) | 213/215 |
| J9 | 0.649 (3) | 0.594 (4) | 0.654 (3) | 0.839 (1) | 0.749 (2) | 0.645 (5) | 0.663 (4) | 0.695 (3) | 0.843 (1) | 0.772 (2) | 126/134 |
| J10 | 0.879 (4) | 0.892 (3) | 0.865 (5) | 0.947 (2) | 0.972 (1) | 0.883 (3) | 0.881 (3) | 0.836 (4) | 0.929 (2) | 0.948 (1) | 1103/1161 |
| avg. | 0.681 | 0.626 | 0.547 | 0.787 | 0.778 | 0.711 | 0.683 | 0.602 | 0.852 | 0.841 | - |

### 6.3 Does the approach to generating insufficient test suites affect the conclusions of RQ2?

In RQ2, we investigate whether the conclusion in RQ1 would change under insufficient test suites. In the literature, many studies simulate insufficient test sets by changing the size of test sets [24, 35-37]. In RQ2, we take a similar idea to generate insufficient test suites. Specifically, we "delete test cases in RQ1 with even index for each program" to simulate an insufficient test suite, which enables readers to reproduce our experiment without randomness. However, for a SUT with the original test suite T, such a generation approach leads to only one specific insufficient test suite. It is unknown whether and to what extent it will affect the conclusions of RQ2. To tackle this problem, in this section, we randomly generate 100 subsets from T to simulate multiple insufficient test suites, each of which contains half of the test cases in T. Table 9 shows the average OP and *VMS* of 100 subsets on 10 programs. From Table 9, we have the following observations:

- The conclusions are consistent with RQ2: OP is a better indicator than *VMS* in mutation reduction evaluation. SMS and CMS are better than RMS, Sentinel, and COS.
- The *VMS* is larger than that in RQ2 and RQ1. This phenomenon reveals that the conclusions of the existing work (i.e. many strategies were reported having a small *VMS* under sufficient test suites and hence were regarded as effective in mutation reduction) worked only on a sufficient test suite. As a result, the practicability of existing work in real application scenarios is questionable.

To conclude, the approach to generating insufficient test suites has little effect on the conclusions of RQ2.

Table 9. The discrimination ability comparison between OP and Variation of MS under multiple insufficient test suites (the strategy rank by the Scott-Knott ESD test is shown in brackets and a lower rank is better)

| Project | OP | | | | | Variation of MS | | | | |
|---|---|---|---|---|---|---|---|---|---|---|
| | RMS | Sentinel | COS | SMS | CMS | RMS | Sentinel | COS | SMS | CMS |
| J1 | 0.642 (3) | 0.410 (4) | - | 0.895 (1) | 0.768 (2) | 0.145 (3) | 0.155 (3) | - | 0.091 (1) | 0.115 (2) |
| J2 | 0.776 (3) | 0.687 (4) | - | 0.903 (1) | 0.810 (2) | 0.076 (1) | 0.121 (2) | - | 0.133 (3) | 0.068 (1) |
| J3 | 0.404 (3) | 0.407 (3) | 0.392 (3) | 0.825 (1) | 0.587 (2) | 0.199 (4) | 0.176 (3) | 0.178 (3) | 0.161 (2) | 0.148 (1) |
| J4 | 0.825 (3) | 0.678 (4) | - | 0.864 (2) | 0.898 (1) | 0.076 (2) | 0.062 (1) | - | 0.058 (1) | 0.074 (2) |
| J5 | 0.615 (3) | 0.507 (4) | 0.476 (5) | 0.665 (1) | 0.636 (2) | 0.102 (1) | 0.193 (4) | 0.298 (5) | 0.160 (3) | 0.133 (2) |
| J6 | 0.544 (3) | 0.483 (5) | 0.509 (4) | 0.727 (1) | 0.715 (2) | 0.119 (2) | 0.184 (3) | 0.080 (1) | 0.300 (4) | 0.189 (3) |
| J7 | 0.575 (3) | 0.655 (2) | 0.556 (4) | 0.709 (1) | 0.663 (2) | 0.180 (3) | 0.213 (4) | 0.136 (2) | 0.174 (3) | 0.084 (1) |
| J8 | 0.536 (3) | 0.515 (4) | 0.539 (3) | 0.695 (1) | 0.670 (2) | 0.192 (1) | 0.204 (1) | 0.191 (1) | 0.375 (3) | 0.306 (2) |
| J9 | 0.586 (3) | 0.572 (3) | 0.554 (4) | 0.768 (1) | 0.750 (2) | 0.135 (2) | 0.106 (1) | 0.142 (2) | 0.284 (3) | 0.127 (2) |
| J10 | 0.735 (4) | 0.755 (3) | 0.730 (4) | 0.893 (1) | 0.798 (2) | 0.057 (1) | 0.078 (2) | 0.053 (1) | 0.182 (3) | 0.081 (3) |
| avg. | 0.624 | 0.567 | 0.537 | 0.794 | 0.730 | 0.128 | 0.149 | 0.154 | 0.192 | 0.133 |

### 6.4 What is the possible reason that Sentinel has a low OP value?

From the experimental results, we can find that Sentinel has a low OP value. Since Sentinel is a state-of-the-art work on mutant reduction, we want to figure out the possible reasons why it has an unsatisfactory OP value. In our opinion, the reasons for this are mainly from two-fold. First, Sentinel is originally proposed for cross-version prediction (rather than for cross-project prediction), i.e., it is trained on the old version of a project and is tested on the new version of the same project [8]. In their experiment, in most case, there is only 0%~3% code churn when comparing the test version with the training version. In other words, the test version is very similar to the training version. As a result, it is understandable that the reduction strategy learned by Sentinel will have a good performance on the test version. However, in our context, we train Sentinel on one program and test it on another program. In nature, this is a cross-project (program) prediction scenario, where the training and test programs are very different. This is one possible reason why Sentinel does not perform well in our study. Second, the optimization objectives and the used basic strategies in Sentinel determine that it is hard to achieve a high OP value. As introduced in Section 4.2.2, Sentinel has two optimization objectives: the average "absolute" strategy effectiveness and the execution time, which are not directly related to the "order-preserving ability". Furthermore, Sentinel uses OP-ineffective basic strategies such as RMS and COS to generate a reduction strategy. Consequently, it is not surprising that Sentinel will lead to a low OP value.

# 7 IMPLICATIONS

Our study has important implications for both researchers and practitioners. By analyzing the pitfalls of existing indicators, we propose two useful indicators to evaluate the effectiveness of a mutant reduction strategy. Furthermore, we apply these two indicators to compare the effectiveness of five reduction strategies. The detailed implications are listed as follows:

- *Our work warns that existing reduction evaluation indicators should be used with caution*. We show that the existing indicators such as *VMS* and $E_s$ have a low ability to distinguish between mutation reduction strategies or even may lead to counter-intuitive conclusions. In particular, for the most commonly used indicator *VMS*, we theoretically prove that a random reduction strategy RMS is the perfect strategy in expectation. This means that, under *VMS*, any pretty reduction strategy at most performs similarly to RMS. The above facts warn that existing indicators are unable to objectively depict the effectiveness of mutation reduction. Therefore, in the future studies, the existing indicators such as *VMS* and $E_s$ should be used and explained with caution, especially when using them to identify a good reduction strategy.

- *Our work provides simple but useful indicators for mutation reduction evaluation*. From the viewpoint of measurement theory, we show that the essential of mutation reduction evaluation is to evaluate the "order-preserving-ability" of a reduction strategy, i.e., to what extent the mutation score order among test suites is maintained before and after mutation reduction. Furthermore, we proposed two simple but useful indicators, OP and EROP, to measure the "order-preserving-ability" of a reduction strategy. In particular, we provide a light-weight "continuous half-sample" computation approach. Consequently, for a mutation reduction strategy, given a SUT with the original test suite and mutation set, OP and EROP can be efficiently and easily computed. As a result, we suggesting using OP and EROP to evaluate mutation reduction effectiveness in practice.

- *Our work discloses the practical value of important mutation reduction strategies*. We apply OP and EROP to compare the effectiveness of five important mutation reduction strategies, including RMS, Sentinel, COS, SMS, and CMS. We find that, when the reduction ratio is less than 50%, in most cases, RMS has a strong "order-preserving-ability" (OP > 0.9) and the other strategies do not exhibit an advantage (EROP ≈ 0). Furthermore, CMS exhibits a similar or higher "order-preserving-ability" than RMS, regardless of which mutation reduction ratio is considered. Therefore, from the viewpoint of practical use, RMS is a good option if the reduction ratio is less than 50% and the mutation reduction cost is considered. However, if the mutation reduction cost is not a concern, SMS and CMS are preferred. In particular, from the viewpoint of academic research, there is a room to develop more effective mutation reduction strategies. The reason is that SMS and CMS have a EROP well below 0.1 in most cases.

# 8 THREATS TO VALIDITY

In this section, we discuss the important threats to the construct validity, internal validity, and external validity of our study. Construct validity denotes the extent to which the variables used in our study accurately measure what we purport to measure. Internal validity is the degree to which conclusions can be drawn about the causal effect of independent variables on the dependent variable. External validity is the degree to which the results of the research can be generalized to the population under study and other research settings.

## 8.1 Construct validity

In our study, we use OP and EROP to measure the effectiveness of a mutation reduction strategy. Ideally, for a SUT with the original test suite T, the population OP and EROP values should be estimated using a "continuous subsample" approach from T. However, due to the high computation complexity, it is hard to apply such an approach in practice. To tackle this problem, we use a "continuous half-sample" approach to replace a "continuous subsample" approach to estimate OP and EROP. Therefore, the most important threat is that a "half-sample" approach may lead to biased OP and EROP values. In order to mitigate this threat, we run k (in default, k = 100) repeated "continuous half-sample" to generate OP and EROP. Indeed, from the viewpoint of statistical theory, subsample and half-sample can produce similar estimates [41, 42]. Therefore, this threat should not have a large influence on the construct validity of OP and EROP.

## 8.2 Internal validity

The first threat is the selection bias of the subject mutation reduction strategies. In RQ1, we compare five mutation reduction strategies in the experiment when investigating the discrimination ability of OP and EROP. The reasons are two-fold. On the one hand, these five reduction strategies are either very popular or the state-of-the-art reduction strategies. On the other hand, the other reduction strategies such as a higher-order mutant strategy may need to introduce new mutants, which violates our experimental principle of "selecting a mutant subset from all mutants" in this study. In the future, we will use more mutation reduction strategies to investigate the discrimination ability of OP and *VMS* to reduce this threat.

The second threat is that the way to generate insufficient test suites may lead to a biased conclusion. In RQ2, for each program, we "delete test cases in RQ1 with even index for each program" to simulate an insufficient test suite. As a result, the generalization of the conclusion under RQ2 may be questionable. To mitigate this threat, we re-run the experiment (see Section 6.3) by randomly generating 100 subsets from the original test set associated with each program. The experimental results show that the conclusion in RQ2 is stable. In this sense, this threat has been minimized as possible as we can.

The third threat is that the specific setting of the step length in reduction ratio may lead to a biased conclusion. In RQ3, we "set the reduction ratio from 95% to 5% with a step length of 5%" to investigate the influence. It is unknown the influence of a different step setting on the conclusion in RQ3. In order to reduce this threat, we used a step length of 1% instead of 5% to rerun the experiment. Consequently, we found that the conclusion in RQ3 remained no change. Therefore, this threat has been considered in our study.

## 8.3 External validity

In this study, we use ten subject programs with a variety of test suites and mutant sets to investigate the effectiveness of OP and EROP. The overall results show that OP and EROP perform better than the existing indicators such as VSM and $E_s$ in mutation reduction evaluation, especially in distinguishing between mutation reduction strategies. This is due to the fact that OP and EROP aim to measure the "order-preserving" ability, rather than the "mutation-score-preserving" ability. Given this situation, it is reasonable to believe that similar findings would be observed if different programs, test suites, and mutation sets are used. Nonetheless, we do not claim that our findings can be generalized to the population. Indeed, this is an inherent problem to most (if not all) empirical studies, not unique to us. To mitigate this threat, there is a need to replicate our study across a wider variety of programs, test suites, and mutation sets in the future work.

## 9 CONCLUSION AND FUTURE WORK

The motivation of this study is to understand the pitfalls of existing mutation reduction evaluation indicators and explore how to objectively measure the ability of a mutation reduction strategy to maintain test suite effectiveness evaluation. In the last decades, many mutation reduction strategies have been proposed and their reduction effectiveness are evaluated by *VMS* and $E_s$. From our analysis and experimental results, however, we can see that *VMS* and $E_s$ may lead to misleading conclusions and have a weak ability to distinguish various mutation reduction strategies. The fundamental reason is that they are not originally designed to directly measure to what extent the mutation score order among test suites is changed before and after mutation reduction. Indeed, from the viewpoint of measurement theory, for mutation reduction evaluation, the essential is to evaluate the "order-preserving ability" of a mutation reduction strategy, which is missing in our community. To this end, we propose two indicators, OP and EROP, for "order-preserving ability" evaluation. OP evaluates the degree to which the reduction strategy maintains a relative evaluation of a test suite, while EROP indicates the effort-aware OP relative to a random reduction strategy. In particular, OP and EROP can be easily and efficiently computed. Our experimental results show that OP and EROP have a better ability to distinguish among various mutation reduction strategies than *VMS* and $E_s$. Furthermore, we find that SMS and CMS are more effective than the other mutant selection strategies under OP and EROP.

In the future, on the one hand, we plan to study how to employ OP and EROP to develop more effective mutation reduction strategies. As shown in our experimental results, the average OP of existing approaches is lower than 0.9, indicating a large room for improvement. This is especially true for Sentinel. As the first step toward this direction, we will hence explore how to improve Sentinel: first, we will modify the basic strategy of Sentinel by adding SMS into the basic strategy set or using CMS instead of grouping mutants; second, we will change the objective function to be OP oriented. On the other hand, we plan to apply OP and EROP to the evaluation of high-order mutant reduction strategies. In our current study, we focus on these mutation reduction strategies without adding new mutants. However, OP and EROP can also be directly used to evaluate high-order mutant selection strategies. Such an evaluation will inform developers whether and which high-order mutant reduction strategies are effective in practice.

## REFERENCES

[1] Y. Jia, M. Harman. An analysis and survey of the development of mutation testing. IEEE Transactions on Software Engineering, 37(5), 2011: 649-678.

[2] M. Papadakis, M. Kintis, J. Zhang, Y. Jia, Y. Traona, M. Harmanb. Mutation testing advances: An analysis and

survey. Advance in Computers, 112, 2019: 275-378.

[3] A. S. Namin, J. H. Andrews, D. J. Murdoch. Sufficient mutation operators for measuring testing effectiveness. ICSE 2008: 10-18.

[4] J. Offutt, A. Lee, G. Rothermel, R. H. Untch, C. Zapf. An experimental determination of sufficient mutant operators. ACM Transactions on Software Engineering and Methodology, 5(2), 1996: 99-118.

[5] P. Delgado-Prez, S. Segura, I. Medina-Bulo. Assessment of C++ object-oriented mutation operators: A selective mutation approach. Software Testing, Verification and Reliability, 27(4-5), 2017: n/a-n/a.

[6] M. E. Delamaro, J. Offutt, P. Ammann. Designing deletion mutation operators. ICST 2014:11-20.

[7] M. E. Delamaro, L. Deng, V. H. S. Durelli, N. Li, J. Offutt. Experimental evaluation of SDL and one-op mutation for C. ICST 2014: 203-212.

[8] G. Guizzo, F. Sarro, J. Krinke, S. Vergilio. Sentinel: A hyper-heuristic for the generation of mutant reduction strategies. IEEE Transactions on Software Engineering, 2020. https://doi.org/10.1109/TSE.2020.3002496

[9] M. Papadakis, N. Malevris. An empirical evaluation of the first and second order mutation testing strategies. ICST 2010:90-99.

[10] B. Kurtz, P. Ammann, J. Offutt, M. E. Delamaro, M Kurtz, N. Gokce. Analyzing the validity of Selective mutation with dominator mutants. FSE 2016:571-582.

[11] D. Gong, G. Zhang, X. Yao, F. Meng. Mutant reduction based on dominance relation for weak mutation testing. Information and Software Technology. 81, 2017: 82-96.

[12] R. A. DeMillo, R. J. Lipton, F. G. Sayward. Hints on test data selection: help for the practicing programmer. IEEE Computer, 11(4), 1978: 34-41

[13] B. Kurtz, P. Ammann, M. E. Delamaro, J. Offutt, L. Deng. Mutant subsumption graphs. ICSTW 2014: 176–185.

[14] R. Just, B. Kurtz, P. Ammann. Inferring mutant utility from program context. ISSTA 2017: 284-294

[15] R. J. Lipton, F. G. Sayward. The status of research on program mutation. In: Proceedings of the Workshop on Software Testing and Test Documentation, 1978: 355-373.

[16] A. P. Mathur, W. E. Wong. An empirical comparison of data flow and mutation-based test adequacy criteria. Software Testing, Verification and Reliability 4,1994:9-31.

[17] W. E. Wong, A. P. Mathur. Reducing the cost of mutation testing：an empirical study. Journal of Systems and Software. 31(3), 1995: 185-196.

[18] B. Kurtz, P. Ammann, J. Offutt. Static analysis of mutant subsumption. ICST 2015: 1-10

[19] L. Zhang, S. Hou, J. Hu, T. Xie, H. Mei. Is operator-based mutant selection superior to random mutant selection? ICSE 2010: 435-444.

[20] R. Gopinath, I. Ahmed, M. A. Alipour, C. Jensen, A. Groce. Do Mutation Reduction Strategies Matter? Technical report, Oregon State University, 2015.

[21] M. Gligoric, A. Groce, C. Zhang, R. Sharma, M. A. Alipour, D. Marinov. Guidelines for coverage-based comparisons of non-adequate test suites. ACM Transactions on Software Engineering and Methodology, 24(4), 2015: 1-33.

[22] M. Yu, Y. Ma: Possibility of cost reduction by mutant clustering according to the clustering scope. Software

Testing, Verification and Reliability, 29(1-2), 2019: n/a-n/a.

[23] A. P. Mathur. Performance, effectiveness, and reliability issues in software testing. COMPSAC,1991:604-605.

[24] S. Hussain. Mutation clustering．M.S. thesis London, UK King's College. 2008.

[25] P. Zhang, Y. Li, W. Ma, Y. Yang, L. Chen, H. Lu, Y. Zhou, B. Xu. CBUA: A probabilistic, predictive, and practical approach for evaluating test suite effectiveness. IEEE Transactions on Software Engineering, 2020. https://doi.org/10.1109/TSE.2020.3010361

[26] J. Zhang, L. Zhang, M. Harman, D. Hao, Y. Jia, L. Zhang, Predictive mutation testing. IEEE Transactions on Software Engineering, 45(9), 2019: 898-918.

[27] R. Gopinath, I. Ahmed, M. A. Alipour, C. Jensen, A. Groce. Mutation Reduction Strategies Considered Harmful. IEEE Transactions on Reliability, 66(3) 2017: 854-874.

[28] Apache. http://commons.apache.org

[29] R. Gopinath, M. A. Alipour, I. Ahmed, C. Jensen, A. Groce. On the limits of mutation reduction strategies. ICSE 2016: 511-522

[30] T. Laurent, M. Papadakis, M. Kintis, C. Henard, Y. L. Traon, A. Ventresque. Assessing and improving the mutation testing practice of pit. ICST 2017: 430–435.

[31] R. Just, D. Jalali, L. Inozemtseva, M. D. Ernst, R. Holmes, G. Fraser. Are mutants a valid substitute for real faults in software testing? FSE 2014: 654-665.

[32] PIT. http://pitest.org/

[33] Cobertura. https://github.com/cobertura

[34] PIT mutators. https://pitest.org/quickstart/mutators

[35] P. S. Kochhar, F. Thung, D. Lo. Code coverage and test suite effectiveness: Empirical study with real bugs in large systems. SANER 2015: 560-564.

[36] Y. T. Chen, R. Gopinath, A. Tadakamalla, M. D. Ernst, R. Holmes, G. Fraser, P. Ammann, R. Just. Revisiting the relationship between fault detection, test adequacy criteria, and test set size. ASE 2020: n/a-n/a.

[37] L. Inozemtseva, R. Holmes. Coverage is not strongly correlated with test suite effectiveness. ICSE 2014: 435–445.

[38] M. Papadakis, C. Henard, M. Harman, Y. Jia, Y. L. Traon. Threats to the validity of mutation-based test assessment. ISSTA 2016:354-365.

[39] C. Tantithamthavorn, S. McIntosh, A.E. Hassan, K. Matsumoto. An empirical comparison of model validation techniques for defect prediction models. IEEE Transactions on Software Engineering, 43(1), 2017: 1-18.

[40] J. Cohen. Statistical power analysis for the behavioral sciences. Routledge. ISBN 978-1-134-74270-7, 1988.

[41] G. Babu. Subsample and half-sample methods. Annals of the Institute of Statistical Mathematics, 44(4), 1992: 703-720.

[42] J. Shao, X. Shi. Half-sample variance estimation. Communications in Statistics: Theory and Methods, 18(11), 1989: 4197-4210.